\documentclass[prl,aps,twocolumn,longbibliography]{revtex4-1}

\usepackage{color}
\usepackage{bbm} 
\usepackage{amsfonts,amsmath,amssymb,stmaryrd}
\usepackage{graphicx}
\usepackage{subfigure}  
\usepackage{hyperref}
\usepackage{epsfig}
\usepackage{mathrsfs}
\usepackage{verbatim}
\usepackage{ulem}	
\usepackage{siunitx}
\usepackage{soul} 
\usepackage[all]{hypcap}
\usepackage{braket}

\newcommand{\hlf}{\mbox{$\frac{1}{2}$}}

\newcommand{\sig}{\hat{\sigma}}

\usepackage{array}

\begin{document}
\normalem	

\title{Mobile bound states of Rydberg excitations in a lattice}

\author{Fabian Letscher}
\affiliation{Department of Physics and Research Center OPTIMAS, 
University of Kaiserslautern, D-67663 Kaiserslautern, Germany}
\affiliation{Graduate School Materials Science in Mainz, 
Gottlieb-Daimler-Strasse 47, D-67663 Kaiserslautern, Germany}

\author{David Petrosyan}
\affiliation{Institute of Electronic Structure and Laser, FORTH, 
GR-71110 Heraklion, Crete, Greece}

\begin{abstract}
Spin lattice models play central role in the studies of 
quantum magnetism and non-equilibrium dynamics of spin excitations -- magnons. 
We show that a spin lattice with strong nearest-neighbor interactions 
and tunable long-range hopping of excitations can be realized by 
a regular array of laser driven atoms, with an excited Rydberg state 
representing the spin-up state and a Rydberg-dressed ground state 
corresponding to the spin-down state. 
We find exotic interaction-bound states of magnons that propagate 
in the lattice via the combination of resonant two-site hopping 
and non-resonant second-order hopping processes.
Arrays of trapped Rydberg-dressed atoms can thus serve as a flexible platform 
to simulate and study fundamental few-body dynamics in spin lattices.
\end{abstract}

\date{\today}

\maketitle

\paragraph*{Introduction.}
Interacting many-body quantum systems are notoriously difficult to simulate
on classical computers, due to the exponentially large Hilbert space and 
quantum correlations between the constituents. It was therefore suggested
to simulate quantum physics with quantum computers \cite{Feynman1982}, 
or universal quantum simulators consisting of spin lattices with
tunable interactions between the spins \cite{Lloyd1996}. 
Dynamically controlled spin lattices can realize digital and analog 
quantum simulations.  
Quantum field theories not amenable to perturbative treatments are often 
discretized and mapped onto the lattice models for numerical calculations. 
Spin lattices are fundamental to the studies of many solid state systems,
where the competition between the interaction and kinetic energies
determines such phenomena as magnetism and superconductivity.

Realizing tunable spin lattices in the quantum regime is challenging.
Several systems are being explored to this end, 
including trapped ions \cite{IonTraps2012,Zhang2017},
superconducting circuits \cite{SCcircuits2012,Neill2017}, 
quantum dots \cite{QDs2017} and other solid state systems. 
Cold atoms in optical lattice potentials are accurately described 
by the Hubbard model, representing perhaps the most versatile 
and scalable platform to realize various lattice models \cite{Gross995}.
The Hubbard model for two-state fermions or strongly interacting 
bosons at half filling can implement the lattice spin-1/2 model 
\cite{Kuklov2003,Duan2003}. 
The spin-exchange  interaction then stems from the second-order tunneling 
(superexchange) process \cite{Trotzky295,Chen2011} and the interspin 
Ising interaction can exist for atoms or molecules with static 
magnetic or electric dipole moments \cite{Lahaye2009,JunYe2017}.
These interactions are, however, weak (tens of Hz or less), 
which makes the system vulnerable to thermal effects even at ultra-low 
temperatures of nK \cite{Greif1307,Boll1257,Greiner2017}.

\begin{figure}[t]
  \centerline{\includegraphics[width=0.8\columnwidth]{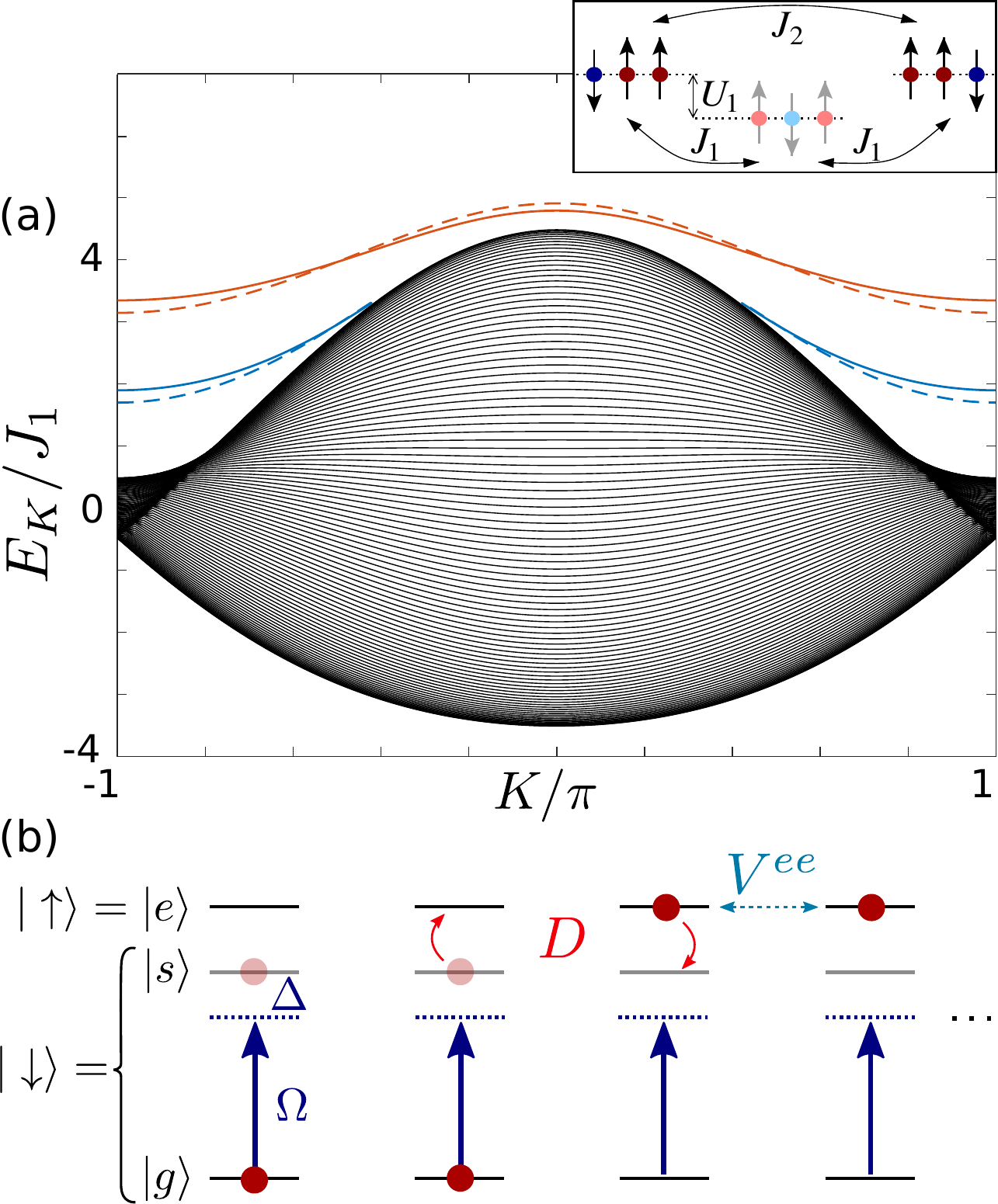}}
  \caption{ (a) Spectrum of two spin (Rydberg) excitations in a lattice 
    versus the center of mass quasi-momentum $K$. 
    The scattering states form a continuum spectrum (black), 
    Eq.~(\ref{eq:EKkscatd}).
    The bound states for strong (red lines) and weak (blue lines) repulsive
    interactions are obtained from the spin-lattice Hamiltonian (dashed lines),
    Eq.~(\ref{eq:BoundPairEnergy}), 
    and from exact diagonalization of the Hamiltonian for the system 
    sketched in (b) (solid lines). In the simulations,
    we used a lattice of size $L=100$ and periodic boundary conditions,
    with the spin model parameters $J_2/J_1=1/8$, $U_1/J_1 = 3.4,1.9$ 
    for the red and blue dashed lines, respectively.
    Inset illustrates the motion of the bound pair via 
    resonant two-site hopping $J_2$ and second order hopping $J_1^2/U_1$. 
    (b)~Level scheme of atoms to realize a spin-lattice model. 
    Atoms in Rydberg states $\ket{e}$ and $\ket{s}$ undergo dipole-dipole
    exchange interaction $\ket{es} \to \ket{se}$ with rate $D$. 
    The atomic ground state $\ket{g}$ is dressed with the Rydberg
    state $\ket{s}$ by a non-resonant laser with Rabi frequency $\Omega$ and 
    detuning $\Delta \gg \Omega$. The spin-up and spin-down states 
    correspond to $\ket{\uparrow} = \ket{e}$ and 
    $\ket{\downarrow} \simeq \ket{g} + \frac{\Omega}{\Delta} \ket{s}$. 
    Interactions $V^{ee}$ between the atoms in state $\ket{e}$ lead to
    formation of mobile bound states of Rydberg excitations. 
    The parameters in numerical simulations shown in (a) correspond 
    to $\Delta/\Omega = 10$, $D_1/\Omega = 1$, $V^{es}_1/\Omega = -0.125$, 
    and $V^{ee}_1/\Omega = 0.03, 0.015$ for the red and blue solid lines, 
    respectively. }
  \label{fig:BoundPairSpectrum}
\end{figure}

Here we propose a practical realization of a tunable spin lattice $XXZ$ 
model with an array of trapped atoms \cite{Barredo1021,Endres2016}. 
The atomic ground state dressed by a non-resonant laser with a Rydberg 
state \cite{Bouchoule2002,Johnson2010,MacriPohl2014} represents 
the spin down state, while another Rydberg state corresponds 
to the spin-up state (see Fig.~\ref{fig:BoundPairSpectrum}). 
Controllable spin-exchange interactions are then mediated by the 
dressing laser and resonant dipole-dipole exchange interaction   
(scaling with distance $r$ as $1/r^3$) between the atoms on the 
Rydberg transition. van der Waals interactions between the 
excited-state atoms (scaling as $1/r^6$) serve as Ising-type 
interaction between the spins 
\cite{Schauss2012,Schauss1455,Labuhn2016,Lienhard2017,Bernien2017}. 
Due to long lifetimes of the Rydberg states and large energy scales 
of their interactions, this system is essentially at zero temperature. 
This permits observation of coherent quantum dynamics of 
spin-excitations -- magnons. 

We study the dynamics of magnons in the spin-lattice with long-range 
spin-excitation hopping  and nearest neighbor interactions. 
Apart form scattering states, we find exotic interaction-bound states of
magnons \cite{Fukuhara2013}. The bound pairs of magnons can propagate 
in the lattice via resonant two-site spin exchange and non-resonant 
second-order exchange interactions [see Fig.~\ref{fig:BoundPairSpectrum}(a)]. 
We note that the spin lattice $XXZ$ model can be mapped onto 
the extended Hubbard model with spinless fermions or hard-core bosons: 
In the extended Hubbard model with low filling, particle tunneling 
from site to site and the attractive or repulsive interactions between 
the particles at the neighboring sites correspond, in the spin-lattice 
model, to the excitation hopping via spin-exchange and to the Ising 
interspin interaction, respectively. 
The bound states of magnons are then equivalent to interaction bound states
of particles in the (extended) Hubbard model \cite{Winkler2006, Fukuhara2013}.
But our solution goes beyond the bound-state solutions of the Hubbard model 
\cite{Piil2007, Petrosyan2007a, Valiente2008, Valiente2009, Valiente2010} 
and it can be easily generalized to arbitrary-range hopping and interactions. 
We find that longer-range hopping of individual magnons leads to the
increased, and tunable, mobility of the bound pairs of magnons.

\paragraph*{Interacting spin excitations in a lattice.}
%
We consider a spin lattice model described by Hamiltonian ($\hbar = 1$)
\begin{equation}
\label{eq:SpinModel}
\mathcal{H} = \sum_{i<j} J_{ij} \left( \sig^+_i \sig^-_j 
+ \sig^-_i \sig^+_j \right) + \sum_{i<j} U_{ij} \hat n_i \hat n_j , 
\end{equation}
where $\sig_i^+ = \ket{\uparrow}_i\bra{\downarrow}$ and 
$\sig_i^- = \ket{\downarrow}_i\bra{\uparrow}$ are the raising 
and lowering operators for the spin at position $i$, and 
$\hat n_i \equiv \sig_i^+ \sig_i^- = \ket{\uparrow}_i \bra \uparrow$
is the projector onto the spin-up state. In Eq.~(\ref{eq:SpinModel}), 
the first term is responsible for the spin transport via the exchange 
interaction $J_{ij}$, while the second term describes the interaction 
between the spins in state $\ket{\uparrow}$ with strength $U_{ij}$.
Both $J_{ij}$ and $U_{ij}$ have finite range and depend only 
on the distance $r=|i-j|$ between the spins at positions $i$ and $j$. 

%
Hamiltonian (\ref{eq:SpinModel}) preserves the number of spin excitations. 
For a single excitation, the interaction does not play a role, 
and the Hamiltonian reduces to 
$\mathcal{H}^{(1)}_J = \sum_{x=1}^L \sum_{d \geq 1} J_d (\ket{x} \bra{x+d} 
+ \ket{x} \bra{x-d} )$, 
where 
$\ket{x} \equiv \sig^+_x \ket{\downarrow_1 \downarrow_2 \ldots \downarrow_L}$
denotes the state with the spin-up at position $x$ in a lattice of $L \gg 1$ 
spins (we assume periodic boundary conditions), and $d=1,2,\ldots$ is 
the range of the exchange interaction. The transformation 
$\ket{x} = \frac{1}{\sqrt{L}} \sum_q e^{i q x} \ket{q}$ diagonalizes 
the Hamiltonian,
$\mathcal{H}^{(1)}_J = \sum_q  \ket{q} \bra{q} E^{(1)}_q$,
which indicates that the plane waves 
$\ket{q} = \frac{1}{\sqrt{L}} \sum_x e^{i q x} \ket{x}$
with the lattice quasi-momenta $q = \frac{2\pi \nu}{L}$ 
($\nu = -\frac{L-1}{2}, \ldots , \frac{L-1}{2}$) 
are the eigenstates of $\mathcal{H}^{(1)}_J$ with the eigenenergies
$E^{(1)}_q = \sum_{d \geq 1} 2 J_d \cos(q d)$.

%
Consider now two spin excitations. We denote by $\ket{x,y}$ the state with 
one spin-up at position $x$ and the second spin-up at $y > x$. 
With this notation, the transport and interaction terms of the 
Hamiltonian~(\ref{eq:SpinModel}) are given by
\begin{align}
\label{eq:2ExcitationHoppingHamiltonian}
\mathcal{H}^{(2)}_J = \sum_{x<y} & 
\Big[ \sum_{d} \, J_d (\ket{x,y}\bra{x-d,y} + \ket{x,y}\bra{x,y+d}) \nonumber \\
+ & \sum_{d < y-x} \!\! J_d (\ket{x,y}\bra{x+d,y} + \ket{x,y}\bra{x,y-d}) \nonumber \\
+ & \sum_{d > y-x} \!\! J_d (\ket{x,y}\bra{y,x+d} + \ket{x,y}\bra{y-d,x}) \Big] ,
\\
\label{eq:2ExcitationInteractionHamiltonian}
& \mathcal{H}^{(2)}_U = \sum_{x<y} U_{xy} \ket{x,y}\bra{x,y} .
\end{align}
We introduce the center of mass $R \equiv (x+y)/2$ and 
relative $r\equiv y-x$ coordinates. Making the transformation 
$\ket{R} = \frac{1}{\sqrt{\tilde L}} \sum_K e^{iKR} \ket{K}$ 
($\tilde L = 2L -3$), we obtain the total Hamiltonian 
$\mathcal{H}^{(2)} = \mathcal{H}^{(2)}_J + \mathcal{H}^{(2)}_U$
that is diagonal in the basis $\ket{K}$ of the center of mass 
quasi-momentum $K = \frac{2\pi \nu}{\tilde{L}}$:
$\mathcal{H}^{(2)} = \sum_K  \ket{K} \bra{K} \otimes \mathcal{H}_K$,
where
\begin{align}
\mathcal{H}_K = \sum_{r} & \Big[ \sum_{d} J_{d,K} \ket{r}\bra{r+d} 
+ \sum_{d < r} J_{d,K} \ket{r}\bra{r-d} 
\nonumber \\ & 
+ \sum_{d > r} J_{d,K} \ket{r}\bra{d-r} + U_{r} \ket{r}\bra{r} \Big] ,
\label{eq:HtotK}
\end{align}
with $J_{d,K} \equiv 2 J_d \cos(Kd/2)$ \cite{SM}. 
The two-body wavefunction can be cast as 
$\ket{\Psi(x,y)} = \frac{1}{\sqrt{\tilde L}} \sum_K e^{iKR} \ket{K} 
\otimes \sum_{r \geq 1} \psi_K (r) \ket{r}$,
where the relative coordinate wavefunction $\psi_K (r)$ 
depends on the quasi-momentum $K$ as a parameter via 
the effective hopping rates $J_{d,K}$ in $\mathcal{H}_K$.
There are two kinds of solutions of the eigenvalue problem
$\mathcal{H}_K \ket{\psi_K} = E_K \ket{\psi_K}$
for $\ket{\psi_K} = \sum_{r \geq 1} \psi_K(r) \ket{r}$, 
corresponding to scattering states of asymptotically free magnons and 
to the interaction-bound states. 

%
The wavefunction for the scattering states has the standard 
form containing the incoming and scattered plane waves 
$\psi_{K,k} (r > d_U) = e^{ikr} + e^{-2 i \delta_{K,k}} e^{-ikr}$, 
where $d_U$ is the (finite) range of the interaction potential $U_r$, and 
the phase shift $\delta_{K,k}$ depends on $U_r$. The energies of the scattering 
states are simply given by the sum of energies of two free magnons,
\begin{equation}
E_{K,k}^{(\mathrm{s})} = E^{(1)}_{q_1} + E^{(1)}_{q_2} = \sum_d 2 J_{d,K} \cos (kd) ,
\label{eq:EKkscatd}
\end{equation}
where $K = q_1 +q_2$ and $k= (q_1-q_2)/2$ are the center of mass and relative 
quasi-momenta. 
In Fig.~\ref{fig:BoundPairSpectrum}(a) we show the spectrum of the scattering
states, assuming the range of the spin-exchange interaction $d_{J} =2$ 
with $J_1 > J_2$, while $J_{d \geq 3} = 0$. Note that due to the longer 
range hopping $J_{2}$, the spectrum at $K=\pm \pi$ does not reduce to a 
single point $E^{(\mathrm{s})} =0$ as in \cite{Valiente2008, Valiente2009}, 
but has a finite width $E_{K=\pi,k}^{(\mathrm{s})} \in [-4J_2,4J_2]$, see also 
\cite{Piil2007}. 

%
The bound state solutions correspond to a normalizable relative coordinate 
wavefunction, $\sum_r |\psi_K(r)|^2 =1$. 
We assume nearest-neighbor interaction, 
$U_1 \neq 0$ and $U_{r>1} = 0$ in Eq.~(\ref{eq:HtotK}).
We set $\psi_K(0) = 0$ and $\psi_K(1) = c$, with $c$ some constant,
and make the ansatz
\begin{equation}
\psi_K(r) = \alpha_K \psi_K(r-1) + \beta_K \psi_K(r-2) . \label{eq:psirrecrel}
\end{equation}
The physical intuition behind this recurrence relation is that every  
(discrete) position $r$ can be reached from positions $r-1$ and $r-2$ 
with the amplitudes $\alpha_K \propto J_1$ and $\beta_K \propto J_2$.
We then obtain \cite{SM} 
$\alpha_K = \frac{J_{1,K}}{U_1}$, $\beta_K = \frac{J_{2,K}}{U_1+J_{2,K}}$,
and the energy of the bound state 
\begin{align}
\label{eq:BoundPairEnergy}
E_K^{(\mathrm{b})} =&  2 J_{2,K} + \frac{J_{1,K}^2}{U_1}  
+ \frac{J_{1,K}^2 J_{2,K}}{U_1^2} + \frac{U_1^2}{U_1 + J_{2,K}} .
\end{align}
The first term on the right-hand-side of this equation does not depend 
on the interaction $U_1$ and it describes two-site resonant hopping of 
the excitation over the other excitation, 
$\ket{x-1,x} \leftrightarrow \ket{x,x+1}$, with rate $\propto J_2$.
This process is resonant because the relative distance $r=1$, and thereby the 
interaction energy, are conserved during this two-excitation ``somersault''. 
The second and third terms are contributions from the second-order 
($\propto J_1^2/U_1$) and third-order ($\propto J_1^2 J_2/U_1^2$) hopping 
processes. The last term is the energy shift due to interaction $U_1$. 

\begin{figure}[t]
  \centerline{\includegraphics[width=0.8\columnwidth]{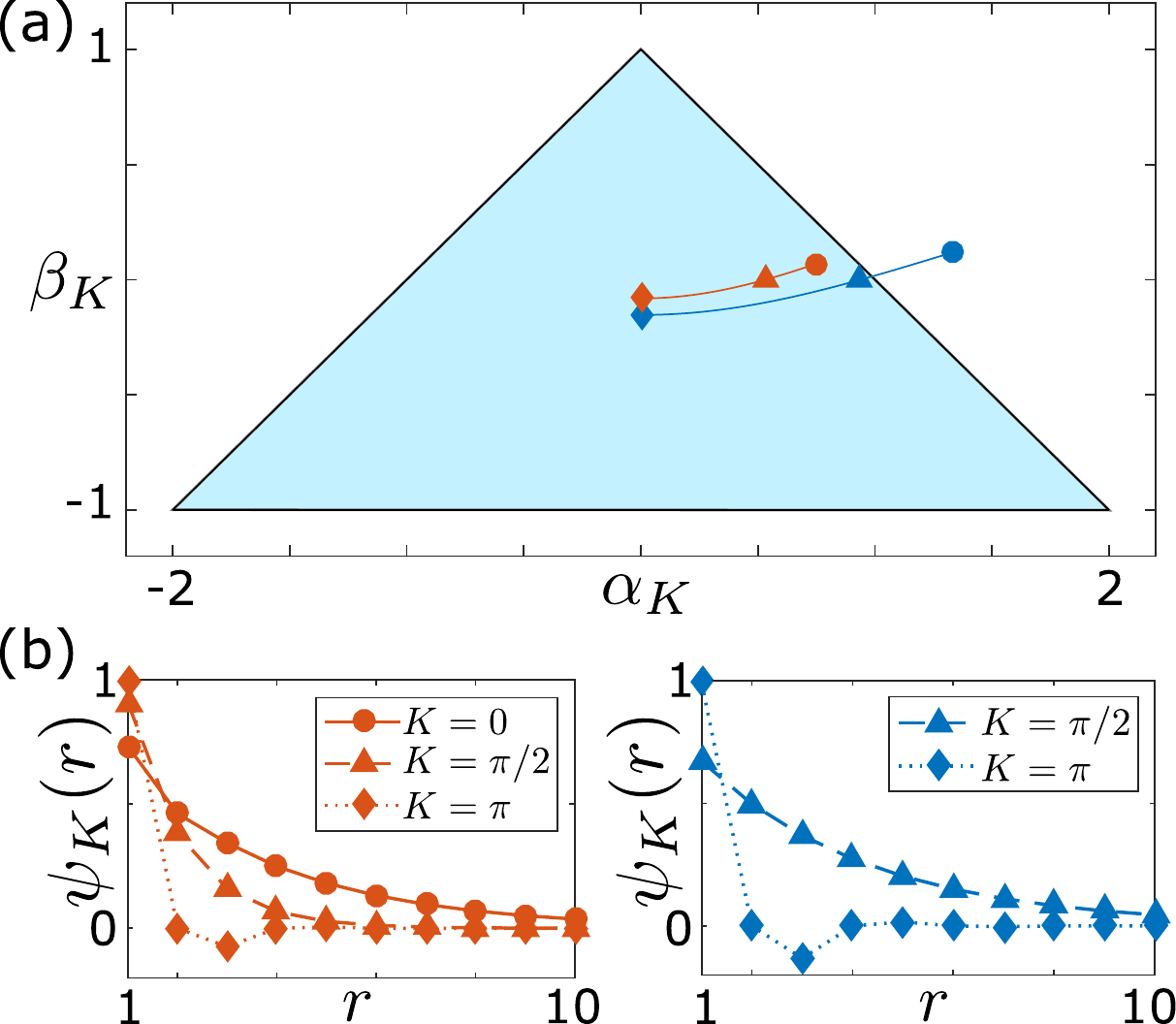}}
  \caption{(a) Diagram of values of $\alpha_K$, $\beta_K$ for the existence 
    of bound states (light-blue shaded region).
    (b) Wavefunction $\psi_K(r)$ versus the relative distance $r$ 
    for several values of the center of mass quasi-momentum $K$. 
    The parameters are the same as in Fig. \ref{fig:BoundPairSpectrum}, 
    with $U_1/J_1 =3.4$ (left graph and red line in (a)), and 
    $U_1/J_1 =1.9$ (right graph and blue line in (a)) where 
    the bound state does not exist in the vicinity of $K=0$.}
  \label{fig:BoundWavefunctionCriteria}
\end{figure}

\begin{figure}[b]
  \centerline{\includegraphics[width=0.8\columnwidth]{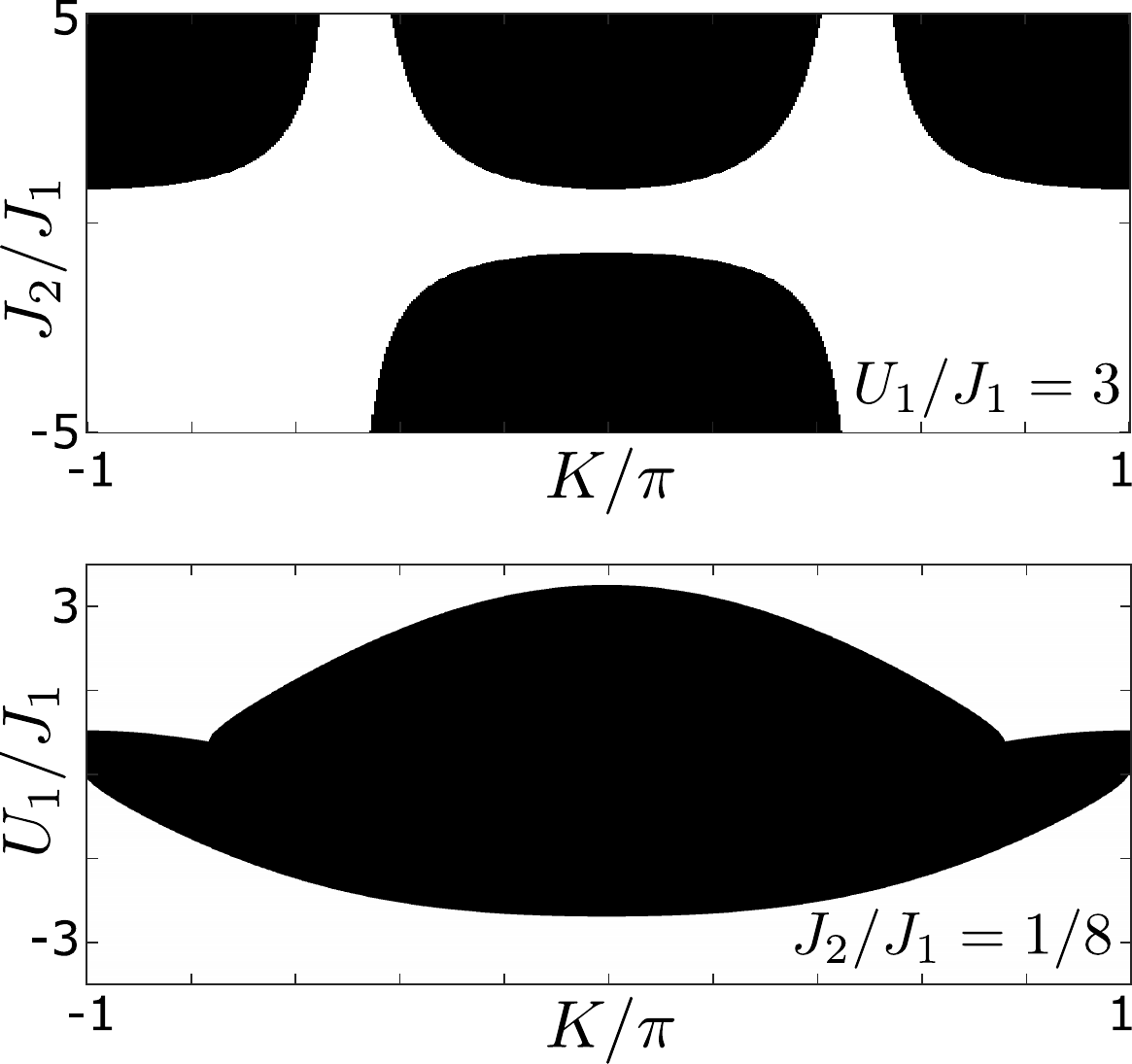}}
  \caption{Diagram of values of $J_2/J_1$, for fixed $U_1=3 J_1$ (upper panel),
    and $U_1/J_1$, for fixed $J_2=J_1/8$ (lower panel), versus $K$, for the 
    existence (white regions) and absence (black regions) of the bound states.}
  \label{fig:ExistenceOfBoundStates}
\end{figure}

The above solution is valid under the conditions that bound-state 
wavefunction is normalizable. Inserting $\psi_K(r) \propto \lambda^r$ 
into Eq.~(\ref{eq:psirrecrel}), we obtain that the wavefunction  
exponentially decays with distance $r$, and therefore is normalizable, 
when $\hlf |\alpha_K \pm \sqrt{\alpha_K^2 + 4 \beta_K^2}| < 1$.
In Fig.~\ref{fig:BoundWavefunctionCriteria}(a) we show the values of $\alpha_K$ 
and $\beta_K$, forming a triangular region, for which there exists 
an exponentially localized bound state. With only nearest-neighbor 
hopping ($J_2 = 0$), we recover the condition $|\alpha_K| < 1$ of 
Refs.~\cite{Valiente2008,Valiente2009}. 
For a given set of parameters $J_1,J_2,U_1$, the bound state may not exists 
for all values of the center of mass quasi-momentum $K$, since both 
$\alpha_K$ and $\beta_K$ depend on $K$. In general, the closer is the 
point ($\alpha_K,\beta_K$) to the boundary of the shaded region
in Fig.~\ref{fig:BoundWavefunctionCriteria}(a), the less localized 
is the bound state wavefunction, as we illustrate with two examples 
in Fig.~\ref{fig:BoundWavefunctionCriteria}(b).  
In Fig.~\ref{fig:ExistenceOfBoundStates} we show the diagrams of $J_2/J_1$ 
and $U_1/J_1$ versus $K$ for the existence of the bound states. 
Clearly, for certain sets of parameters, the bound states do not exist at all, 
or exist only within a certain interval of values of $K$. 

\paragraph*{Rydberg dressed atoms in a lattice.}
The spin lattice model of Eq.~(\ref{eq:SpinModel}) might be realized 
with a regular array of atoms in Rydberg states $\ket{s}$ and $\ket{e}$. 
We could excite one or more atoms to state 
$\ket{e} = \ket{\uparrow}$ and prepare all the remaining atoms 
in state $\ket{s} = \ket{\downarrow}$. Assuming the transition 
$\ket{s} \leftrightarrow \ket{e}$ is dipole allowed, 
resonant dipole-dipole interaction between the atoms separated 
by $d$ lattice sites would lead to transfer of excitations via the 
exchange interaction $\ket{es} \leftrightarrow \ket{se}$ with rate 
$D_{d} = \frac{C_3/a^3}{d^3}$, where $C_3$ is the interaction coefficient 
and $a$ is the lattice constant \cite{Barredo2015,DeLeseleuc2017}.
Atoms in the Rydberg states also interact via the van der Waals interactions
$V^{\mu \nu}_{d} = \frac{C_6^{\mu \nu}/a^6}{d^6}$, which would map onto the 
interactions $U$ between the spin excitations 
\cite{Schauss2012,Schauss1455,Labuhn2016,Lienhard2017,Bernien2017}, 
provided $V^{ee}$ differs from 
the interaction $V^{es}$ between the $\ket{e}$ and $\ket{s}$ state atoms.

Typically, however, the resonant dipole-dipole interaction $D$ is  
orders of magnitude stronger than the van der Waals interactions $V$, 
since the latter originate from non-resonant dipole-dipole interactions, 
$V \sim D^2/\delta \omega$, with large F\"orster defects, 
$\delta \omega \gg D$ \cite{Saffman2010}. Small interactions $V \ll D$ 
will preclude the interplay between the spin transport and spin-spin 
interactions. To mitigate this problem, 
we propose to dress trapped ground state atoms with the Rydberg state
$\ket{s}$. The dressing laser would then mediate hopping of the Rydberg 
excitation $\ket{e}$ to nearby atoms in the dressed ground state with 
rates $J_d$ which can be made comparable to, or even weaker than, 
the effective interaction $U_r$ between the excitations. 
Rydberg dressing of ground-state atoms 
\cite{Bouchoule2002,Henkel2010,Johnson2010,Pupillo2010,Wuster2011,MacriPohl2014} 
is a versatile tool for tuning interatomic interactions to simulate 
various lattice models 
\cite{Schempp2015,Glaetzle2015,vanBijnen2015,Buchmann2016,Zeiher2016c,Zeiher2017}. 

We consider an array of single atoms with the level scheme 
shown in Fig. \ref{fig:BoundPairSpectrum}(b).
The ground state $\ket{g}$ of each atom is coupled to the Rydberg 
state $\ket{s}$ by a laser with Rabi frequency $\Omega$ and large 
detuning $\Delta \gg \Omega$. We assume that $\Delta$ is much larger 
than the resonant dipole-dipole interactions $D_{d}$ between the $\ket{s}$ 
and $\ket{e}$ state atoms separated by $d=1,2,\ldots$ lattice sites. 
The van der Waals interactions $V^{\mu \nu}_{r}$ are assumed 
to be still weaker, so the hierarchy of the energy scales
is $\Delta \gg \Omega, D_{d} \gtrsim V^{ee}_{r}, V^{es}_{r},V^{ss}_{r}$.

The laser instills a small admixture $\frac{\Omega}{\Delta} \ket{s}$ 
of the Rydberg state to the ground state $\ket{g}$ \cite{SM}. 
We then identify the dressed ground state with the spin-down state, 
$\ket{\downarrow} \simeq \ket{g} + \frac{\Omega}{\Delta} \ket{s}$,
while the spin-up state is $\ket{\uparrow} = \ket{e}$. 
Neglecting the interactions $\sim \frac{\Omega^4}{\Delta^4} V^{ss}_{r}$ between 
the dressed ground state atoms, we adiabatically eliminate the nonresonant 
state $\ket{s}$ and obtain effective excitation hopping rates 
$J_{d} \simeq \frac{\Omega^2 D_{d}}{\Delta^2}$ 
between the atoms separated by $d$ lattice sites. Since $D_d \propto d^{-3}$,
we can truncate $J_{d}$ to range $d_J = 2$. More careful considerations 
show that the hopping rates $J_{1,2}$ for a Rydberg excitation are slightly 
altered when another Rydberg excitation is in a close proximity \cite{SM}. 
We assume that the lifetime of the Rydberg state $\ket{e}$ is longer than 
the timescale $J_d^{-1}$ for the system dynamics and neglect dissipation. 
The number of atoms prepared in state $\ket{e}$ is then conserved. 
Decay via the non-resonant state $\ket{s}$ is suppressed 
by the factor of $\frac{\Omega^2}{\Delta^2}$.

For the effective interaction potential between the excitations we obtain
$U_r \simeq V^{ee}_r + 2 \frac{\Omega^2 D_r^2}{\Delta^3}$, 
where both terms scale with distance as $\propto r^{-6}$. 
We assume that $U_r$ is dominated by the nearest-neighbor van der Waals 
interaction $V^{ee}_1$ between the atoms in Rydberg states $\ket{e}$. 
Corrections to the level shift of Rydberg dressed atoms in the vicinity 
of the Rydberg excited atom $\ket{e}$ lead to small contribution to
$U_r$ and weak longer range interaction \cite{SM}.
Despite these small variations of $J_{d}(r)$ and $U_r$ with distance
$r$ between Rydberg excitations, the spin-lattice model approximates 
well the properties of interacting Rydberg excitations, including the 
two-excitation bound states shown in Fig.~\ref{fig:BoundPairSpectrum}(a). 

%
The dynamics of Rydberg excitations in a lattice and their bound states
can be prepared and observed with the presently available experimental 
techniques. We envisage an array of single atoms confined in a chain 
of microtraps \cite{Barredo1021,Endres2016}. 
Using focused laser beams, selected atoms can be resonantly excited 
from the ground state $\ket{g}$ to the Rydberg state $\ket{e}$, 
while the dressing laser is turned off, $\Omega =0$. 
Next, turning on the dressing laser, $\Omega \neq 0$ will lead to the
admixture of the Rydberg state $\ket{s}$ to the ground state atoms, which
will induce the $\ket{e}$ excitation hopping between the atoms due the
dipole-dipole exchange interaction. With realistic experimental 
parameters \cite{SM}, hopping rates $J_1 \simeq 200\:$kHz and $J_2 = J_1/8$ 
can be achieved. This will allow observation of non-trivial dynamics 
of the excitations on the timescale of Rydberg state lifetimes 
$\tau \gtrsim 100\:\mu$s.
With a proper choice of state $\ket{e}$, we can ensure appropriate
interaction strength $U_1 \simeq V^{ee} \gtrsim J_1$,
which will result in the formation of tightly bound Rydberg excitations 
that are still mobile as they propagate with rate $\sim J_2$. 
Free Rydberg excitations and their scattering states can be 
discriminated from the interaction-bound states spectroscopically 
or by the fast and slow dynamics, respectively.  
Turning off the dressing laser would freeze the dynamics and individual 
Rydberg excitations can be detected with high efficiency and 
single-site resolution \cite{Labuhn2016,Bernien2017,Lienhard2017}. 

\paragraph*{Conclusions.}

We have shown that spin lattice models with controllable 
long range hopping and interactions between the spin excitations can 
be realized with Rydberg dressed atoms in a lattice. We have found 
mobile bound states of spin excitations which are quantum lattice solitons. 
It would be interesting to consider bound aggregates of more than two magnons
which may form mobile clusters that can propagate via resonant long-range
hopping process. In turn, multiple clusters can form a lattice liquid  
\cite{Mattioli2013,Dalmonte2015}, while including controllable dephasing 
and disorder \cite{Schempp2015,Schonleber2015} may change the transport 
of (bound) Rydberg excitations from ballistic to diffusive or localized. 
Hence, this system can be used to simulate and study few- and many-body 
quantum dynamics in spin lattices. 


\begin{acknowledgments}
We thank Michael Fleischhauer and Manuel Valiente 
for valuable advice and discussions.
F.L. is supported by a fellowship through the Excellence Initiative 
MAINZ (DFG/GSC 266) and by DFG through SFB/TR49. 
D.P. is supported in part by the EU H2020 FET Proactive project RySQ.
We are grateful to the Alexander von Humboldt Foundation for travel 
support via the Research Group Linkage Programme. 
\end{acknowledgments}


\bibliography{MBS.bib}


\newpage


\section{Supplemental Material}


\subsection{Details of derivation of the two-excitation wavefunction 
in a spin lattice}
\label{ap:BoundState}

Consider two spin excitations in a lattice. The transport and interaction
Hamiltonians are given by Eqs.~(\ref{eq:2ExcitationHoppingHamiltonian})
and (\ref{eq:2ExcitationInteractionHamiltonian}) in the main text, namely
\begin{align}
\label{eq:app:2ExcitationHoppingHamiltonian}
\mathcal{H}^{(2)}_J = \sum_{x<y} & 
\Big[ \sum_{d} \, J_d (\ket{x,y}\bra{x-d,y} + \ket{x,y}\bra{x,y+d}) \nonumber \\
+ & \sum_{d < y-x} \!\! J_d (\ket{x,y}\bra{x+d,y} + \ket{x,y}\bra{x,y-d}) \nonumber \\
+ & \sum_{d > y-x} \!\! J_d (\ket{x,y}\bra{y,x+d} + \ket{x,y}\bra{y-d,x}) \Big] ,
\end{align}
and
\begin{align}
\label{eq:app:2ExcitationInteractionHamiltonian}
\mathcal{H}^{(2)}_U = \sum_{x<y} U_{xy} \ket{x,y}\bra{x,y} ,
\end{align}
where $\ket{x,y}$ denotes the state with the excited spins at positions $x$ 
and $y > x$. 

We introduce the center of mass $R \equiv (x+y)/2$ and relative $r\equiv y-x$ 
coordinates: 
$R = 1+\hlf, 2, 2+\hlf, \ldots , L-\hlf$ takes $\tilde L = 2L -3$ 
discrete values, and $r = 1,2,\ldots,L-1$ takes $L-1$ values. 
In terms of these coordinates, the transport Hamiltonian reads
\begin{widetext}
\begin{align}
\label{eq:app:2ExcitationRr}
\mathcal{H}^{(2)}_J = \sum_{R,r} & 
\Big[ \sum_{d} J_d (\ket{R}\bra{R-d/2} \otimes \ket{r}\bra{r+d} 
+ \ket{R}\bra{R+d/2} \otimes \ket{r}\bra{r+d} ) \nonumber \\
+ & \; \sum_{d < r} J_d (\ket{R}\bra{R+d/2} \otimes \ket{r}\bra{r-d} 
+ \ket{R}\bra{R-d/2} \otimes \ket{r}\bra{r-d}  ) \nonumber \\
+ & \; \sum_{d > r} J_d (\ket{R}\bra{R+d/2} \otimes \ket{r}\bra{d-r}
+ \ket{R}\bra{R-d/2} \otimes \ket{r}\bra{d-r}) \Big] ,
\end{align}
\end{widetext}
Similarly to the single excitation case, we can diagonalize the center 
of mass part of $\mathcal{H}^{(2)}_J$ by the transformation 
$\ket{R} = \frac{1}{\sqrt{\tilde L}} \sum_K e^{iKR} \ket{K}$, where 
$K = \frac{2\pi \nu}{\tilde{L}}$ ($\nu = -\frac{\tilde{L}-1}{2}, \ldots , 
\frac{\tilde{L}-1}{2}$) is the center of mass quasi-momentum:
\begin{align}
\label{eq:app:2ExcitationKr}
\mathcal{H}^{(2)}_J = \sum_{K} \ket{K}\bra{K} \otimes \sum_{r}  &
\Big[ \sum_{d} J_{d,K} \ket{r}\bra{r+d} \nonumber \\
+ & \; \sum_{d < r} J_{d,K} \ket{r}\bra{r-d} \nonumber \\
+ & \; \sum_{d > r} J_{d,K} \ket{r}\bra{d-r} \Big] ,
\end{align}
where $J_{d,K} \equiv 2 J_d \cos(Kd/2)$. The interaction Hamiltonian 
remains diagonal in these coordinates,
\begin{align}
\label{eq:app:2HU}
\mathcal{H}^{(2)}_U = \sum_{K} \ket{K}\bra{K} \otimes \sum_{r} U_{r} \ket{r}\bra{r} ,
\end{align}
and the total Hamiltonian can be cast as
\[
\mathcal{H}^{(2)} = \mathcal{H}^{(2)}_J + \mathcal{H}^{(2)}_U = 
\sum_K  \ket{K} \bra{K} \otimes \mathcal{H}_K.
\]
We have thus reduced the two-body problem for  
\begin{align}
\ket{\Psi(x,y)} & = \sum_{x<y}  \Psi(x,y) \ket{x,y}
\nonumber \\
& = \frac{1}{\sqrt{\tilde L}} \sum_K e^{iKR} \ket{K} \otimes \sum_{r \geq 1} \psi_K (r) \ket{r} 
\end{align} 
to a one-body problem for the relative coordinate wavefunction $\psi_K (r)$,
which depends on the center of mass quasi-momentum $K$ as a parameter via 
the effective hopping rates $J_{d,K}$ in $\mathcal{H}_K$.

Our aim is to solve the eigenvalue problem
\begin{equation}
\mathcal{H}_K \ket{\psi_K} = E_K \ket{\psi_K}  \label{eq:app:eigenproblemHk}
\end{equation}
for the relative coordinate wavefunction 
$\ket{\psi_K} = \sum_{r \geq 1} \psi_K(r) \ket{r}$. 
The scattering solutions are expressed via the plane waves 
as given in the main text. We present here the details of derivation 
of the bound solutions corresponding to a normalizable [localized] 
relative coordinate wavefunction, $\sum_r |\psi_K(r)|^2 =1$
[with $\psi_K(r\to \infty) \to 0$]. 

We assume range $d_{U} =1$ (nearest-neighbor) interaction, $U_1 \neq 0$ 
and $U_{r>1} = 0$ in Eq.~(\ref{eq:app:2HU}), leading to the Hamiltonian 
\begin{align}
\label{eq:app:HK}
\mathcal{H}_K = \sum_{r\geq 1} & 
\Big[ J_{1,K} (\ket{r} \bra{r+1} + \ket{r+1}\bra{r} ) \nonumber \\
+ & \; J_{2,K} (\ket{r} \bra{r+2} + \ket{r+2}\bra{r} ) \Big] \nonumber \\
& + (U_1 + J_{2,K}) \ket{1}\bra{1} . 
\end{align}
This results in the equation 
\begin{align}
\label{eq:app:eigenproblem}
& J_{1,K} \big[ \psi_K(r+1) + \psi_K(r-1) \big] \nonumber \\
+ & J_{2,K} \big[ \psi_K(r+2) + \psi_K(r-2) \big] \nonumber \\
+ & (U_1 + J_{2,K}) \delta_{r,1} \psi_K(r) = E_K \psi_K(r) . 
\end{align}
We set $\psi_K(0) = 0$ and $\psi_K(1) = c$, with $c$ some constant to
be determined by the normalization. We make an ansatz for the wavefunction,
\begin{equation}
\psi_K(r) = \alpha_K \psi_K(r-1) + \beta_K \psi_K(r-2) . 
\label{eq:app:psirrecrel}
\end{equation}
The physical meaning of this recurrence relation is that every site 
$r$ can be reached from the previous two sites $r-1$ and $r-2$ with the 
amplitudes $\alpha_K \propto J_1$ and $\beta_K \propto J_2$. Starting 
from position $r=1$, the wavefunction at any $r$ can then be written as
\begin{equation}
\label{eq:app:psiransatz}
\psi_K(r) = c \sum_{n=0}^{\lfloor (r-1)/2 \rfloor}
{{r-1-2n}\choose{n}} \alpha_K^{r-1-2n}\beta_K^{n} ,
\end{equation}
where $\lfloor \cdot \rfloor$ is the floor function, and the binomial 
coefficients count the weights for different path from site $1$ to $r > 1$. 
For instance, we can reach $\ket{r=4}$ from $\ket{1}$ by three one-site 
hoppings $\propto \alpha_K^3$, or by two-site hopping $\beta_K$ followed 
by one-site hopping $\alpha_K$, or vice versa, 
$\propto \beta_K \alpha_K + \alpha_K\beta_K = 2 \alpha_K \beta_K$. 
Using the ansatz (\ref{eq:app:psiransatz}) in Eqs.~(\ref{eq:app:eigenproblem}) 
for $r=1,2,3$ we obtain a set of three equations, 
\begin{subequations}
\begin{align}
& E_K = (U_1+J_{2,K}) + J_{1,K} \alpha_K + J_{2,K} (\alpha_K^2+\beta_K), \\
& E_K \alpha_K = J_{1,K} (\alpha_K^2 +\beta_K + 1) + J_{2,K} (\alpha_K^3+2\alpha_K\beta_K), \\
& E_K (\alpha_K^2 + \beta_K)  = J_{1,K} (\alpha_K + \alpha_K^3 + 2\alpha_K \beta_K)\nonumber \\
& \qquad \qquad \qquad \;\; + J_{2,K} (1 + \alpha_K^4 +3\alpha_K^2\beta_K + \beta_K^2) , 
\end{align}
\end{subequations}
for the unknowns $\alpha_K,\beta_K,E_K$. Solving these equations, we obtain 
\begin{align}
\label{eq:app:Coefficients}
\alpha_K = \frac{J_{1,K}}{U_1}, \quad \beta_K = \frac{J_{2,K}}{U_1+J_{2,K}},
\end{align}
while the energy of the bound state is
 \begin{align}
\label{eq:app:BoundPairEnergy}
E_K^{(\mathrm{b})} =&  2 J_{2,K} + \frac{J_{1,K}^2}{U_1}  
+ \frac{J_{1,K}^2 J_{2,K}}{U_1^2} + \frac{U_1^2}{U_1 + J_{2,K}} .
\end{align}
The physical meanings of the various terms of this equation are discussed
in the main text.

We finally discuss the conditions of validity of the above solution 
under which the bound-state wavefunction is normalizable,
$\sum_r |\psi_K(r)|^2 =1$. 
Assuming $\psi_K(r) \propto \lambda^r$ and inserting into 
Eq.~(\ref{eq:app:psirrecrel}), we obtain the quadratic equation 
$\lambda^2 = \alpha_K \lambda + \beta_K$ with the solutions 
\[
\lambda_{1,2} = \frac{\alpha_K \pm \sqrt{\alpha_K^2 + 4 \beta_K^2}}{2} .
\] 
We can now write the wavefunction as 
\begin{equation}
\psi_K(r) = c_1 \lambda_{1}^r + c_2 \lambda_{2}^r ,
\end{equation}
and determine the coefficients $c_{1,2}$ from $\psi_K(0) =0$ and $\psi_K(1) =c$,
leading to 
$c_2 = - c_1 = \frac{c}{\sqrt{\alpha_K^2 + 4 \beta_K^2}}$.
This is of course the same wavefunction as in Eq.~(\ref{eq:app:psiransatz}).
More important, however, is that we have found that 
$\psi_K(r) \propto \lambda_{1,2}^r$ exponentially decays with distance $r$, 
and therefore is normalizable, when both 
$|\lambda_{1,2}| =  \hlf |\alpha_K \pm \sqrt{\alpha_K^2 + 4 \beta_K^2}| < 1$.

\subsubsection{Truncation of interaction range}

\begin{figure}[t]
  \centerline{\includegraphics[width=0.85\columnwidth]{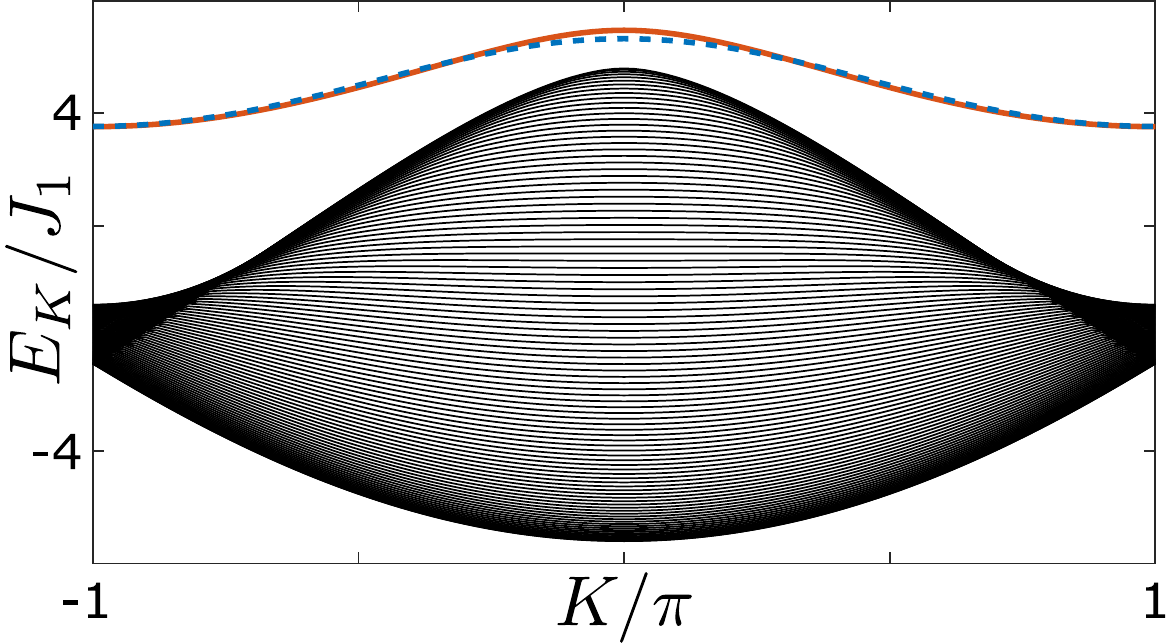}}
  \caption{Scattering (black) and bound-state (red solid line) spectra 
    obtained by exact numerical diagonalization of Hamiltonian with 
    long range interactions $J_d = J_1/d^3$ and $U_r = U_1/r^6$ ($U_1/J_1 = 4$). 
    The bound state energy $E_K$ of Eq.~(\ref{eq:app:BoundPairEnergy}) 
    (dashed blue line), obtained with truncated interactions ($d_J=2$ 
    and $d_U=1$), is nearly indistinguishable from the exact result.}
  \label{fig:TruncatedSpectrum}
\end{figure}

Our formalism to obtain the bound states of excitations in a lattice can
be easily extended to longer range hopping $J_d$ and interaction $U_r$. 
We are, however, mainly concerned with the typical case of resonant 
dipole-dipole exchange interaction, leading to $J_{d} \propto 1/d^3$, 
and van der Waals repulsive or attractive interaction, leading to 
$U_{r} \propto 1/r^6$. We have therefore truncated $J_{d}$ to range $d_J =2$ 
and $U_r$ to range $d_U =1$. In Fig. \ref{fig:TruncatedSpectrum} we show 
the spectra for the scattering and bound states obtained without 
the truncation. This figure clearly demonstrates that the above 
approximations are well justified for the power-law decay of 
the strengths of $J_{d}$ and $U_{r}$ with distance.


\newpage

\subsection{Derivation of the effective excitation hopping rate and 
interaction strength for Rydberg dressed atoms in a lattice}
\label{ap:EffectiveHamiltonian}

Consider an ensemble of atoms in a lattice with period $a$, with one
atom per site. A spatially uniform laser field of frequency $\omega$ 
couples the ground state $\ket{g}$ of each atom to the Rydberg state 
$\ket{s}$ with the Rabi frequency $\Omega$ and detuning 
$\Delta = \omega_{sg} - \omega$, 
see Fig. \ref{fig:BoundPairSpectrum}(b) of the main text.
Resonant dipole-dipole interaction between the atoms at positions 
$i$ and $j$ leads to the exchange interaction 
$\ket{es} \leftrightarrow \ket{se}$ with rate $D_{ij} = \frac{C_3/a^3}{|i-j|^3}$,
where $C_3$ is the interaction coefficient. 
Including also the van der Waals interactions 
$V^{\mu \nu}_{ij} = \frac{C_6^{\mu \nu}/a^6}{|i-j|^6}$ between the Rydberg states,
the Hamiltonian in the rotating frame reads ($\hbar = 1$)
\begin{align}
\label{eq:DressingModel}
\mathcal{H}_{\mathrm{at}} = & \sum_j \left[ \Delta \sig^{ss}_j 
- \Omega \left( \sig^{gs}_j + \sig^{sg}_j \right) \right] 
\nonumber \\ &
+ \sum_{i < j} D_{ij} \left( \sig^{es}_i \sig^{se}_j 
+ \sig^{se}_i \sig^{es}_j \right)
\nonumber \\ & 
+ \sum_{i<j} \left( V^{ee}_{ij} \sig^{ee}_i \sig^{ee}_j 
+ V^{ss}_{ij} \sig^{ss}_i \sig^{ss}_j
+ V^{es}_{ij} \sig^{ee}_i \sig^{ss}_j \right) ,
\end{align}
where $\sig^{\mu \nu}_j = \ket{\mu}_j\bra{\nu}$ are the atomic operators.
%

We take the detuning $\Delta$ of the laser field to be much larger than the 
Rabi frequency $\Omega$ as well as the resonant dipole-dipole interactions 
$D_{d} = \frac{C_3/a^3}{d^3}$ between the Rydberg-state atoms separated 
by $d=1,2,\ldots$ lattice sites. 
The van der Waals interactions $V^{\mu \nu}_{d} = \frac{C_6^{\mu \nu}/a^6}{d^6}$ 
are assumed to be still weaker, 
$\Delta \gg \Omega, D_{d} > V^{ee}_{d}, V^{es}_{d},V^{ss}_{d}$.

\subsubsection{Rydberg dressing}

For a single (isolated) atom, the dipole-dipole and van der Waals
interactions are irrelevant, and the Hamiltonian reduces to that
for a two level system, 
\begin{equation}
\mathcal{H}_{\mathrm{TLS}}
= \Delta \ket{s}\bra{s} - \Omega ( \ket{g}\bra{s} + \ket{s}\bra{s}) .
\end{equation} 
[We set the energy of the ground state $\ket{g}$ to zero and work in 
a rotating frame in which the energy of state $\ket{e}$ is also zero]. 
The eigenstates and corresponding eigenvalues of this Hamiltonian are
\begin{equation}
\ket{\pm} = \frac{\varepsilon_{\mp} \ket{g} + \Omega \ket{s}}
{\sqrt{\varepsilon_{\mp}^2 + \Omega^2}} , \quad 
\varepsilon_{\pm} = \frac{\Delta \pm \sqrt{\Delta^2 + 4 \Omega^2}}{2} .
\end{equation}
For $\Delta \gg \Omega$, the eigenstate 
$\ket{-} \simeq \ket{g} + \frac{\Omega}{\Delta} \ket{s}$, with shifted 
energy $\varepsilon_{-} \simeq - \frac{\Omega^2}{\Delta} \equiv \delta$ 
(ac Stark shift), corresponds to the ground state $\ket{g}$ with 
a small admixture of the Rydberg state $\ket{s}$. We identify 
this Rydberg dressed ground state with the spin-down state, 
$\ket{\downarrow} \equiv \ket{-}$,
while the spin-up state is $\ket{\uparrow}  \equiv \ket{e}$.

A pair of dressed ground-state atoms would interact with each other via 
the Rydberg state $\ket{s}$ components. Each atom is in state $\ket{s}$ with 
probability $\frac{\Omega^2}{\Delta^2}$ and therefore the two-atom interaction 
strength is $\frac{\Omega^4}{\Delta^4} V^{ss}_{r}$ \cite{Buchmann2016}. 
We neglect this weak interaction and instead focus below 
on the interatomic interactions that are up to second order 
in $\frac{\Omega}{\Delta}$.
Hence, with $L$ atoms in a lattice, all in the dressed ground state,
the total energy shift is 
\begin{equation}
E_0 = \sum_i^L \delta_i = - L \frac{\Omega^2}{\Delta} .
\end{equation}
This constant energy shift can be disregarded by redefining the zero-point
energy, e.g., by absorbing the ac Stark shift into the laser detuning,
$\omega \to \omega + \frac{\Omega^2}{\Delta}$.

\subsubsection{Single excitation}

Assume now that one atom is excited to state $\ket{e}$ while the rest 
of the atoms are in the dressed ground state. Our aim is to derive the 
effective hopping rate of the single Rydberg excitation in the lattice 
and the modification of the ac Stark shifts of the ground state atoms 
in the vicinity of the excited one. We are interested in the interatomic
interactions that are up to second order in $\frac{\Omega}{\Delta}$, 
which thus involve no more that one (virtual) $\ket{s}$ excitation.
It is therefore sufficient to consider the two atom state
\begin{equation}
\ket{\phi} = c_{ge} \ket{ge} + c_{eg} \ket{eg} + c_{se} \ket{se} + c_{es} \ket{es}
\end{equation}
and the corresponding Hamiltonian 
\begin{align}
\mathcal{H}^{(1)}_{\mathrm{at}} =&  
\Delta_y^{(x)} \ket{se}\bra{se} + \Delta_x^{(y)} \ket{es}\bra{es} \nonumber \\ &
- \Omega (\ket{ge}\bra{se} + \ket{eg}\bra{es} + \mathrm{H .c.} ) \nonumber \\ & 
+ D_{xy} ( \ket{se}\bra{es} + \mathrm{H.c.} ) , 
\end{align}
where $x$ and $y$ denote the positions of the two atoms, and 
we defined $\Delta_y^{(x)} \equiv \Delta + V^{se}_{xy} = \Delta_x^{(y)}$.
The equations for the amplitudes $c_{\nu \mu}$ of the state vector $\ket{\phi}$ 
are then 
\begin{subequations}
\begin{align}
i \dot{c}_{ge} &= -\Omega c_{se}, \\
i \dot{c}_{eg} &= -\Omega c_{es}, \\
i \dot{c}_{se} &=  (\Delta + V^{se}_{xy}) c_{se} - \Omega c_{ge} + D_{xy} c_{es}, \\
i \dot{c}_{es} &=  (\Delta + V^{se}_{xy}) c_{es} - \Omega c_{eg} + D_{xy} c_{se}. 
\end{align}
\end{subequations}
We adiabatically eliminate states containing the highly detuned 
Rydberg state $\ket{s}$. To that end, we set $\dot{c}_{se} = 0$ 
and $\dot{c}_{es} = 0$ and solve the last two equations for 
${c}_{se}$ and ${c}_{es}$. Inserting the solution into the first 
two equations, we obtain
\begin{subequations}
\begin{align}
i \dot{c}_{ge} &= - \frac{\Omega^2 (\Delta+ V^{se}_{xy})}{(\Delta+ V^{se}_{xy})^2 
- D_{xy}^2} c_{ge} + \frac{\Omega^2 D_{xy}}{(\Delta+ V^{se}_{xy})^2 - D_{xy}^2} c_{eg} ,
\\
i \dot{c}_{eg} &= - \frac{\Omega^2 (\Delta+ V^{se}_{xy})}{(\Delta+ V^{se}_{xy})^2 
- D_{xy}^2} c_{eg} + \frac{\Omega^2 D_{xy}}{(\Delta+ V^{se}_{xy})^2 - D_{xy}^2} c_{ge}.
\end{align}
\end{subequations}
We can interpret these equations as follows: 
The dressed $\ket{g}$ state atom at position $y$ acquires an energy shift 
\begin{equation}
\delta_y^{(x)} = - \frac{\Omega^2 (\Delta+ V^{se}_{xy})}
{(\Delta+ V^{se}_{xy})^2 - D_{xy}^2} ,
\end{equation}
which depends on the position $x$ of the $\ket{e}$ excitation. 
Besides, states $\ket{eg}$ and $\ket{ge}$ are coupled via exchange 
interaction 
\begin{equation}
J_{xy} = \frac{\Omega^2 D_{xy}}{(\Delta+ V^{se}_{xy})^2 - D_{xy}^2} .
\end{equation}
This effective excitation hopping rate $J_{xy} = J_d$ depends on 
the relative distance $d=|x-y|$. 

Hence, the total energy of $L$ atoms in a lattice with a single $\ket{e}$
excitation is 
\begin{equation}
E_1 = \sum_{y \neq x} \delta_y^{(x)} .
\end{equation}
This sum has now $L-1$ terms. The terms $\delta_y^{(x)}$ with small separation 
$|x-y| \geq 1$ are affected by the $D_{xy}$ and $V^{se}_{xy}$ interactions,
while the terms with $|x-y| \gg 1$ are obviously equal to the  
ac Stark shift $\delta = - \frac{\Omega^2}{\Delta}$ of a non-interacting atom. 
Due to the translational invariance of the lattice, $E_1$ does not depend 
on the position $x$ of the $\ket{e}$ excitation. $E_1$ is therefore 
a constant which can be disregarded by redefining the zero-point energy 
[notice, however, that $E_1 \neq E_0$]. 

We thus obtain an effective Hamiltonian for a single excitation hopping 
on a lattice,
\begin{align}
\mathcal{H}^{(1)}_J &= \sum_{x \neq y} J_{xy} \ket{x} \bra{y} 
\nonumber \\ 
&= \sum_{x=1}^L \sum_{d \geq 1} J_d (\ket{x} \bra{x+d} 
+ \ket{x} \bra{x-d} ) , 
\end{align}
which has the same form as $\mathcal{H}^{(1)}_J$ in the main text. 
For $\Delta \gg D_{d}, V^{es}_{d}$, the excitation hopping rates
\begin{equation}
J_{d} \simeq \frac{\Omega^2 D_{d}}{\Delta^2} \propto 1/d^3 
\end{equation}
can be truncated to range $d_J=2$.

\subsubsection{Two excitations}

Consider finally two $\ket{e}$ excitations in the lattice. 
As argued above, to determine interatomic interactions that are up to 
second order in $\frac{\Omega}{\Delta}$, we can restrict our analysis 
to the multiatom configurations with at most one atom in state $\ket{s}$. 
It is then sufficient to consider the three atom state  
\begin{align}
\ket{\phi} =  & \; c_{gee} \ket{gee} + c_{ege} \ket{ege} + c_{eeg} \ket{eeg} 
\nonumber \\
& + c_{see} \ket{see} + c_{ese} \ket{ese} + c_{ees} \ket{ees} .
\end{align}
We assume, as before, that the interaction $V^{ee}_d$ between the $\ket{e}$ 
excitations is weak, $V^{ee}_d \ll \Omega,D_d \ll \Delta$, and neglect it here; 
later we account for $V^{ee}_d$ exactly in the effective Hamiltonian. 
The three-atom Hamiltonian is
\begin{align}
\label{eq:Full3SiteHamiltonian}
\mathcal{H}^{(2)}_{\mathrm{at}} =&
\Delta_{x,y}^{(z)} \ket{ees}\bra{ees} + \Delta_{x,z}^{(y)} \ket{ese}\bra{ese} 
+ \Delta_{y,z}^{(x)} \ket{see}\bra{s ee}
\nonumber \\ & 
- \Omega (\ket{gee}\bra{see} + \ket{ege}\bra{ese} + \ket{eeg}\bra{ees} + 
\mathrm{H.c.} )
\nonumber \\ & 
+ D_{xy} ( \ket{see}\bra{ese} + \mathrm{H.c.} ) 
\nonumber \\ & 
+ D_{xz} ( \ket{see}\bra{ees} + \mathrm{H.c.} )
\nonumber \\ & 
+ D_{yz} ( \ket{ese}\bra{ees} + \mathrm{H.c.} ) ,
\end{align}
where $x,y,z$ denote the positions of the atoms,
$\Delta_{x,y}^{(z)} \equiv \Delta + V^{se}_{xz} + V^{se}_{yz}$ and
similarly for $\Delta_{x,z}^{(y)}$ and $\Delta_{y,z}^{(x)}$.
From the differential equations for the amplitudes $c_{\lambda \mu \nu}$
of $\ket{\phi}$, we adiabatically eliminate the amplitudes 
corresponding to the highly-detuned $\ket{s}$ state, i.e., 
we set $\dot{c}_{ees} = \dot{c}_{ese} = \dot{c}_{see} = 0$, solve
for the amplitudes $c_{ees}, c_{ese},c_{see}$ and insert them 
into the remaining equations. The resulting equations have the form 
\begin{align}
\label{eq:dotc_eeg}
\dot c_{eeg} =& 
\frac{\Omega^2 (\Delta_{x,z}^{(y)} \Delta_{y,z}^{(x)} - D_{xy}^2)} 
{\Gamma(x,y,z)} c_{eeg} 
\nonumber \\ & 
+ \frac{\Omega^2 ( D_{xy} D_{xz} - D_{yz} \Delta_{y,z}^{(x)} )}
{\Gamma(x,y,z)} c_{ege} 
\nonumber \\ & 
+ \frac{\Omega^2 ( D_{xy} D_{yz} - D_{yz} \Delta_{x,z}^{(y)} )}
{\Gamma(x,y,z)} c_{gee},
\end{align}
with 
$\Gamma(x,y,z) \equiv - \Delta_{x,y}^{(z)} \Delta_{x,z}^{(y)} \Delta_{y,z}^{(x)} 
- 2 D_{xy} D_{xz} D_{yz} + \Delta_{x,y}^{(z)} D_{xy}^2 + \Delta_{x,z}^{(y)} D_{x,z}^2 
+ \Delta_{y,z}^{(x)} D_{y,z}^2$, and similarly for $\dot c_{ege}$ 
and  $\dot c_{gee}$.
The first term in Eq.~(\ref{eq:dotc_eeg}) corresponds to the energy shift 
of the dressed $\ket{g}$ state atom, while the other two terms describe 
the exchange interactions between the atom in state $\ket{g}$ and the 
atoms in state $\ket{e}$. 

\begin{figure}[t]
  \centerline{\includegraphics[width=0.7\columnwidth]{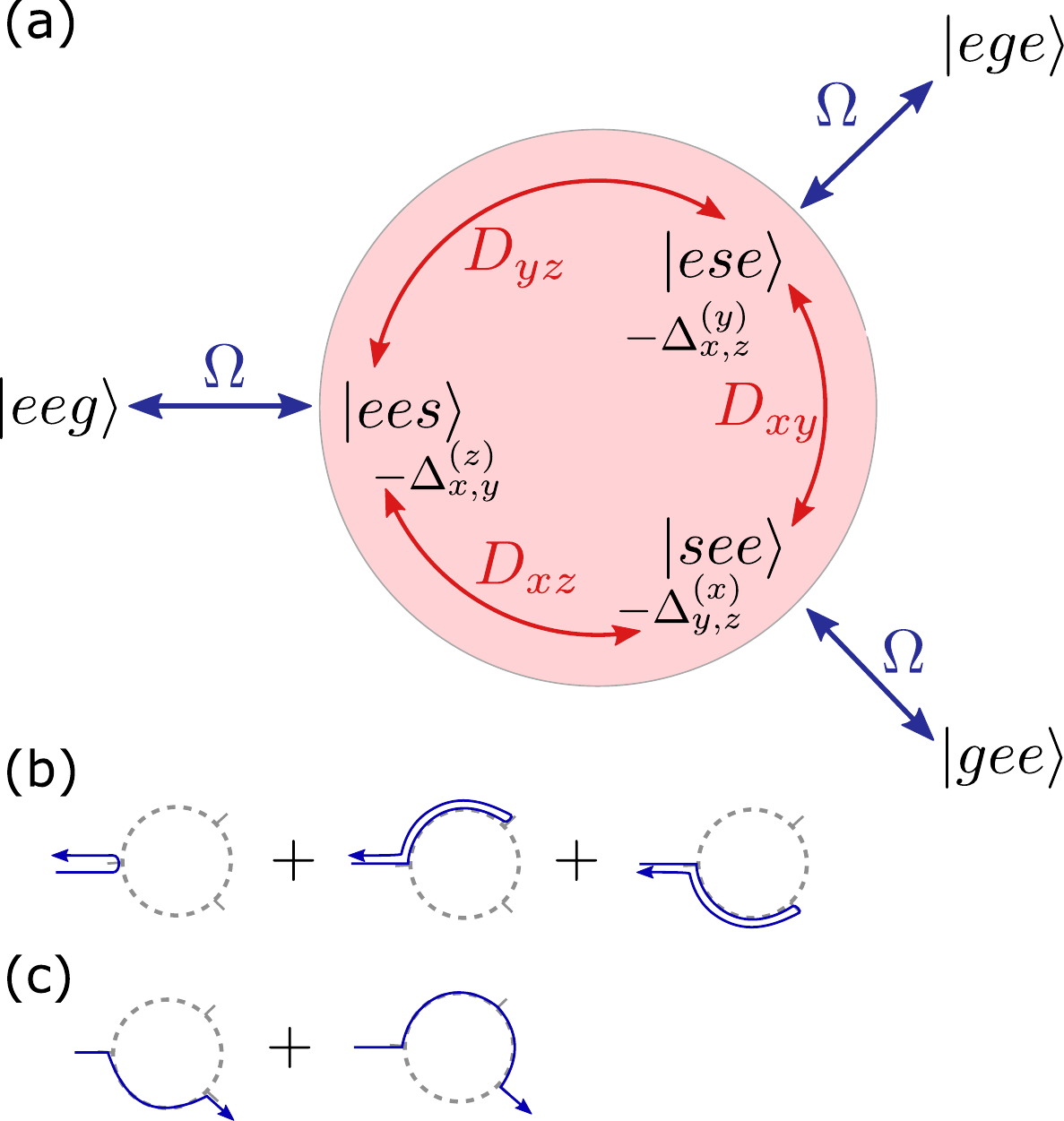}}
  \caption{(a) Diagram of transitions for retrieving the perturbative energy 
    shifts and excitation hopping rates for two excited $\ket{e}$ and one 
    ground $\ket{g}$ state atoms. The atomic positions are $x,y,z$.
    Red-shaded region denotes the high energy subspace,
    $\Delta \gg \Omega, D$, which is eliminated adiabatically. 
    (b) Illustration of three virtual processes contributing to
    the energy shift of $\ket{eeg}$, as per Eq.~(\ref{eq:StarkShift}).
    (c) Two possible paths for the hopping process  
    $\ket{eeg} \leftrightarrow \ket{gee}$ given by 
    Eq.~(\ref{eq:HoppingElement}).}
  \label{fig:PertubationScheme}
\end{figure}

Using series expansion in $\frac{\Omega}{\Delta} \ll 1$, the energy 
shift of the ground state atom at position $z$ can be cast as
\begin{equation}
\delta^{(x,y)}_z = - \frac{\Omega^2}{\Delta_{x,y}^{(z)}} 
- \frac{\Omega^2 D_{xz}^2}{(\Delta_{x,y}^{(z)})^2 \Delta_{z,y}^{(x)}}
- \frac{\Omega^2 D_{y z}^2}{(\Delta_{x,y}^{(z)})^2 \Delta_{x,z}^{(y)}} 
+ O \left( \frac{\Omega^4}{\Delta^4} \right) . \label{eq:StarkShift}  
\end{equation}
Here, the first term is the second order ac Stark shift of the $\ket{g}$ 
state atom due to virtual excitation to state $\ket{s}$ via the non-resonant
laser field. The next two terms describe higher-order shifts due to the laser 
excitation followed by exchange interaction with the $\ket{e}$ state atoms. 
Similarly, we can cast the excitation hopping 
$\ket{eeg} \leftrightarrow \ket{gee}$ between the atoms 
at positions $x$ and $z$ as 
\begin{equation}
J_{xz}^{(y)} = \frac{\Omega^2 D_{xz}}{\Delta_{x,y}^z \Delta_{z,y}^x} 
- \frac{\Omega^2 D_{yz} D_{xy}}{\Delta_{x,y}^z \Delta_{y,z}^x \Delta_{z,x}^y}
+ O \left( \frac{\Omega^4}{\Delta^4} \right) . \label{eq:HoppingElement}
\end{equation}
Here, the first term describes the laser-mediated excitation hopping
via direct dipole-dipole exchange interaction between the atoms at
positions $x$ and $z$. The second term describes the excitation hopping
via indirect process that involves, first, exchange interaction 
between the $\ket{e}$ state atom at position $y$ and the virtually
$\ket{s}$ excited atom at $z$, followed by exchange interaction 
between the $\ket{s}$ state atom now at $y$ and the $\ket{e}$ state 
atom at position $x$. Analogously, we obtain the hopping rates for 
$\ket{eeg} \leftrightarrow \ket{ege}$ and $\ket{ege} \leftrightarrow \ket{gee}$.
In Fig.~\ref{fig:PertubationScheme} we illustrate the virtual processes 
that lead the perturbative energy shifts and excitation hoppings.  

\begin{figure}[t]
  \centerline{\includegraphics[width=0.85\columnwidth]{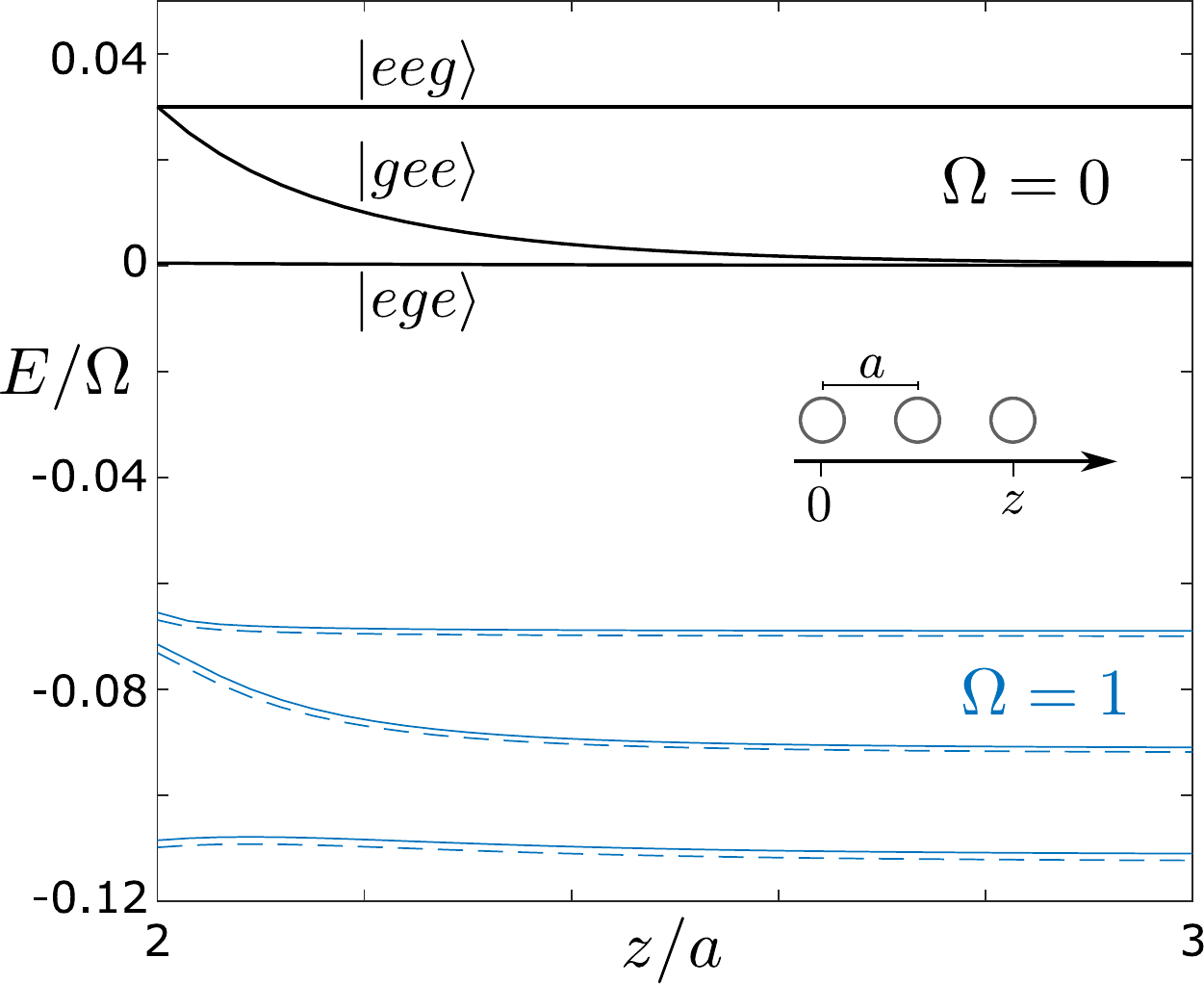}}
  \caption{Comparison of the low energy spectra of the exact Hamiltonian 
    \eqref{eq:Full3SiteHamiltonian} including the interactions $V_r^{ee}$ 
    (solid lines),  
    and the effective Hamiltonian \eqref{eq:Effective3SiteHamiltonian} 
    (dashed lines).
    The positions of the first and second atoms are fixed, $x=0$ and $y=a$, 
    while the position of the third atom vary, $z \geq 2a$. 
    Black lines at $E \geq 0$ show the exact spectrum for $\Omega = 0$, 
    corresponding to the bare states $\ket{ege}$, $\ket{gee}$ and $\ket{eeg}$. 
    Blue lines show the spectra for the dressed states with the parameters
    $\Delta/\Omega = 10$, $D_1/\Omega = 1$, $V_1^{se}/\Omega = -1/8$ and 
    $V_1^{ee}/\Omega = 0.03$ ($D_r \propto 1/r^3$, $V_r \propto 1/r^6$).}
  \label{fig:FullVsEffectiveModel}
\end{figure}

The effective low energy Hamiltonian for two excited and one ground state 
atoms can now be cast as
\begin{align}
\label{eq:Effective3SiteHamiltonian}
\mathcal{H}_{\mathrm{eff}}^{(2)} = & 
  (\delta_{x}^{(y,z)} + V_{yz}^{ee}) \ket{gee}\bra{gee} 
\nonumber \\ &
+ (\delta_{y}^{(x,z)} + V_{xz}^{ee}) \ket{ege}\bra{ege} 
\nonumber \\ &
+ (\delta_{z}^{(x,y)} + V_{xy}^{ee})  \ket{eeg}\bra{eeg}
\nonumber \\ &
+ J^{(z)}_{xy} ( \ket{gee}\bra{ege} + \mathrm{H.c.} )  
\nonumber \\ &
+ J^{(y)}_{xz} ( \ket{eeg}\bra{gee} + \mathrm{H.c.} ) 
\nonumber \\ &
+ J^{(x)}_{yz} ( \ket{ege}\bra{eeg} + \mathrm{H.c.} ) , 
\end{align}
where we have included the interactions $V_r^{ee}$ between the $\ket{e}$
state atoms. In Fig.~\ref{fig:FullVsEffectiveModel} we show the spectrum 
of this Hamiltonian for varying the position $z$ of the third atom, while 
the first and the second atoms are at positions $x=0$ and $y=a$. 
For comparison, we also show the low-energy part of the spectrum 
of the exact Hamiltonian \eqref{eq:Full3SiteHamiltonian} including 
also the interactions $V_r^{ee}$. We observe that the effective Hamiltonian 
reproduces very well the low-energy part of the exact Hamiltonian. 
Clearly, the discrepancy between the exact and effective models 
decreases by increasing the detuning $\Delta$, and in the limit of
$\Omega/\Delta \to 0$ the effective model reduces to the low-energy part 
of the exact model.  

\paragraph{Effective lattice Hamiltonian.}

\begin{figure}[t]
  \centerline{\includegraphics[width=0.85\columnwidth]{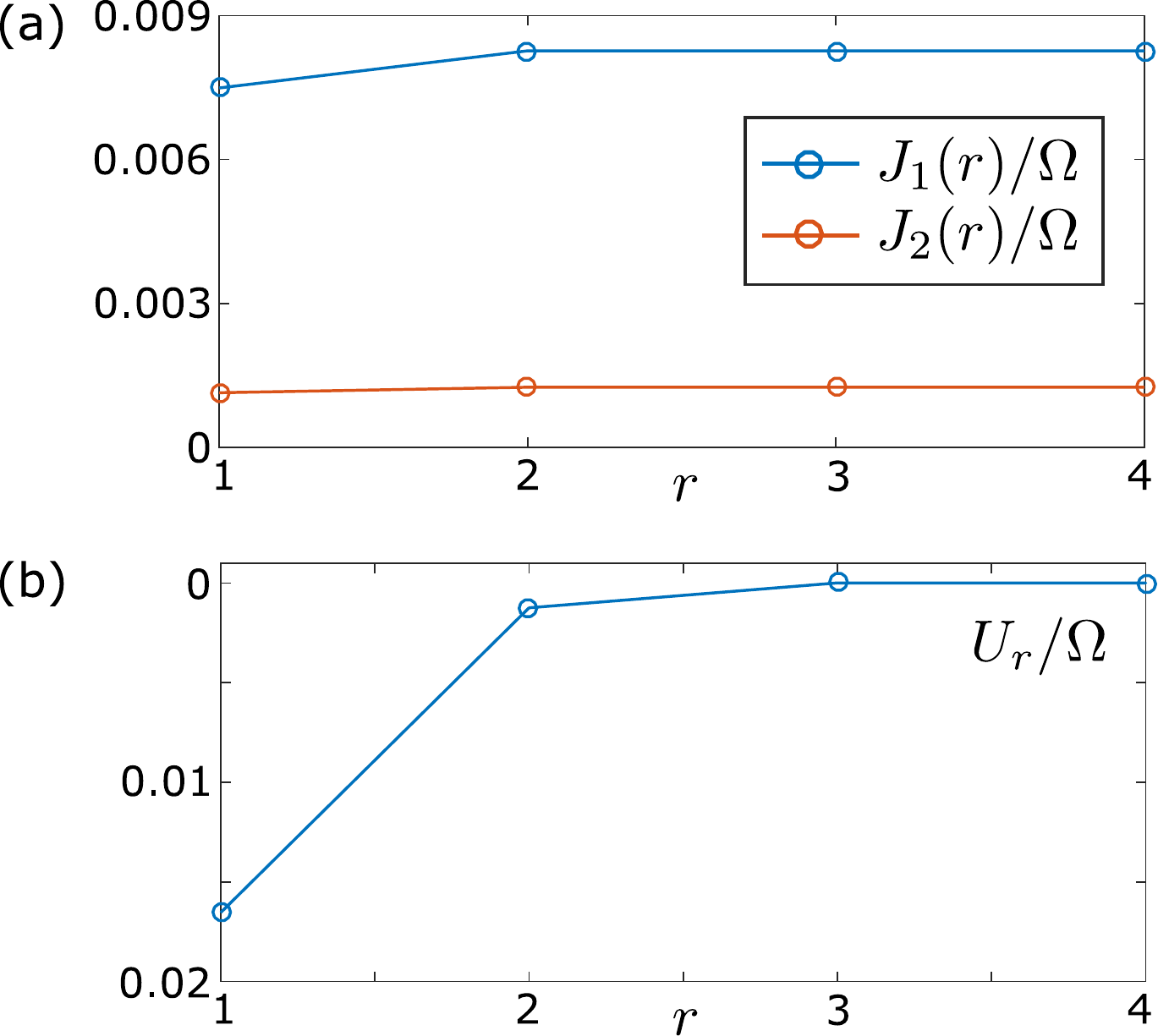}}
  \caption{(a) One- and two-site hopping rates $J_1(r)$ and $J_2(r)$ versus
    distance $r$ between the two excitations. 
    (b)~Interaction potential $U_r$ of Eq.~(\ref{eq:UrAtdef}), for $V^{ee}_1=0$.
    The parameters are $\Delta/\Omega = 10$, $D_1/\Omega = 1$ and 
    $V^{se}_1/\Omega = 1$.  }
  \label{fig:EffJU}
\end{figure}

We can now extend the three atom model to a system of $L$ atoms 
on a lattice (setting the lattice constant $a=1$). 
We start with the transport term of the Hamiltonian. 
Denoting by $x$ and $y$ the positions of the two excitations and 
using the notation $J_{xz}^{(y)} \equiv J_{d}(r)$ with $d \equiv |x-z|$ 
and $r = |x-y|$, we have 
\begin{align}
\label{eq:2AtExcH_J}
\mathcal{H}^{(2)}_J = \sum_{x<y} & 
\Big[ \sum_{d} \, J_d(r) (\ket{x,y}\bra{x-d,y} + \ket{x,y}\bra{x,y+d}) 
\nonumber \\
+ & \sum_{d < y-x} \!\! J_d(r) (\ket{x,y}\bra{x+d,y} + \ket{x,y}\bra{x,y-d}) 
\nonumber \\
+ & \sum_{d > y-x} \!\! J_d(r) (\ket{x,y}\bra{y,x+d} + \ket{x,y}\bra{y-d,x}) 
\Big] ,
\end{align}
which has the same form as Eq.~\eqref{eq:app:2ExcitationHoppingHamiltonian} but 
with the hopping rates $J_d(r)$ that depend on the relative distance $r$ between
the two excitations. Since in the leading order $J_d(r) \propto D_d \sim 1/d^3$,
we truncate it to range $d_J = 2$. 
As in the main text, we can transform $\mathcal{H}^{(2)}_J$ to the center 
of mass $R$ and relative $r$ coordinates and diagonalize the center of mass 
part by Fourier transform 
$\ket{R} = \frac{1}{\sqrt{\tilde L}} \sum_K e^{iKR} \ket{K}$,
obtaining 
\begin{align}
\label{eq:2AtExcKr}
\mathcal{H}^{(2)}_J = & \sum_{K} \ket{K}\bra{K} 
\nonumber \\ & 
\otimes  \Big\{ \sum_{r\geq 1}  
\Big[ 2 J_{1}(r) \cos (K/2) \ket{r} \bra{r+1} + \ket{r+1}\bra{r} ) 
\nonumber \\ & \qquad \quad 
+ 2 J_{2}(r) \cos (K) (\ket{r} \bra{r+2} + \ket{r+2}\bra{r} ) \Big] 
\nonumber \\ &  \qquad \qquad \qquad
+  2 J_{2}'(1) \cos (K) \ket{1}\bra{1} \Big\} .
\end{align}
From Eq. \eqref{eq:HoppingElement} we have for the hopping rates
\begin{subequations} 
\label{eqs:J12r}
\begin{align}
\label{eq:Ja}
J_1(r=1) &= \frac{\Omega^2 D_1}{(\Delta+V^{se}_1)(\Delta+2V^{se}_1)}
\left( 1 - \frac{D_2}{\Delta+V^{se}_1} \right) , \\
J_1(r \geq 2) &= \frac{\Omega^2 D_1}{(\Delta + V^{se}_1 )^2}  , \\
J_2(r=1) &= \frac{\Omega^2 D_2}{\Delta (\Delta+V^{se}_1)} , \\
J_2(r\geq 2) &= \frac{\Omega^2 D_2}{\Delta^2}  , \\
J_2'(1) &= \frac{\Omega^2 D_2}{(\Delta+V^{se}_1)^2} 
\left( 1 - \frac{D_1^2/D_2}{\Delta+2 V^{se}_1}\right) , \label{eq:J2prime}
\end{align}
\end{subequations}
where we set $V^{se}_{d \geq 2} =0$ and $D_{d \geq 3} = 0$. 
In Fig. \ref{fig:EffJU}(a) we show the dependence of the one- and 
two-site hopping rates on the relative distance $r$ between the excitations. 
While $J_2(r)$ is nearly constant for the relevant parameter regime, 
$J_1(r)$ has a noticeable dip at $r=1$ for large $V^{se}_1 \sim \Omega$.
It follows from Eq.~\eqref{eq:Ja} that $J_1(r)$ becomes $r$-independent 
for $V^{se}_1 = -D_2$. Since we assumed that $D_{d} = \frac{C_3/a^3}{d^3}$ 
and $V^{se}_{d} = \frac{C_6^{se}/a^6}{d^6}$, the required lattice constant is
$a = 2\sqrt[3]{-C_6^{se}/C_3}$ with the interaction coefficients $C_6^{se}$ 
and $C_3$ having opposite sign. 
Notice that $J_2'(1)$ in Eq.~(\ref{eq:J2prime}), responsible for the
two-excitation ``somersault'', can be tuned by $\Delta$ or even made 
to vanish. Thus $J_2'(1) = 0$ for $\Delta+2 V^{se}_1 = D_1^2/D_2$, which,
with $D_2 = D_1/8$ and $V^{se}_1 \ll \Delta$, requires $\Delta \simeq 8 D_1$. 

For $\Delta \gg \Omega, D_{1} \gg D_2, V^{es}_{1}$, the excitation hopping rates 
of Eqs.~(\ref{eqs:J12r}) can be well approximated by $r$-independent rates 
\begin{equation}
J_{d} \simeq \frac{\Omega^2 D_{d}}{\Delta^2} \propto 1/d^3 .
\end{equation}

Consider next the effective interaction between the excitations. 
The total energy of $L$ atoms in a lattice with two $\ket{e}$ excitations is
\begin{equation}
E_2(x,y) = \sum_{z \neq x,y} \delta^{(x,y)}_z .  
\end{equation}
This sum has now $L-2$ terms and it depends on the positions $x$ and $y$ 
of the two excitations as per Eq. \eqref{eq:StarkShift}. Due to translational 
invariance of the lattice, $E_2(r)$ depends only on the relative distance 
$r=|x-y|$.  For large $r$, $E_2(r)$ tends to a constant since each dressed 
ground state atom can have at most one excited atom in its vicinity. 
Setting $E_2(r \to \infty)$ as the zero point energy, we can then define 
the interaction potential between the two excitations as
\begin{equation}
U_r = E_2(r) - E_2(r \to \infty) .  \label{eq:UrAtdef}
\end{equation} 
Setting, as before, $V^{se}_{d \geq 2} =0$ and $D_{d \geq 3} = 0$, we obtain 
an effective interaction potential $U_r$ having range $d_U =3$,
\begin{subequations}
\label{eqs:UrAt}
\begin{align}
U_1 =& V^{ee}_1 - 2 \left( \frac{\Omega^2}{\Delta} 
- \frac{\Omega^2}{\Delta+V^{se}_1} \right)
\nonumber \\
& + 2 \frac{\Omega^2 D_1^2}{(\Delta+V^{se}_1)^2} 
\left( \frac{2}{\Delta +V^{ie}_1} - \frac{1}{\Delta + 2 V^{se}_1} \right) 
\nonumber \\ &
+ 2 \Omega^2 D_2^2 \left( \frac{2}{\Delta^3} 
- \frac{1}{\Delta^2(\Delta+V^{se}_1)} - \frac{1}{(\Delta + V^{se}_1)^3}  \right),
\\
U_2 =& V^{ee}_2 - \left( \frac{\Omega^2}
{\Delta} - \frac{\Omega^2}{\Delta+V^{se}_1} \right)
\nonumber \\ &
+ 2 \frac{\Omega^2 D_1^2}{\Delta+V^{se}_1}
\left( \frac{1}{(\Delta +V^{se}_1)^2}  
- \frac{1}{(\Delta + 2 V^{se}_1)^2} \right)
\nonumber \\
&+ 2 \frac{\Omega^2 D_2^2}{\Delta^3},
\\
U_3 &= V^{ee}_3 + 2 \frac{\Omega^2 D_2^2}{\Delta} 
\left( \frac{1}{\Delta^2} - \frac{1}{(\Delta + V^{se}_1)^2} \right) ,
\end{align}
\end{subequations}
where for consistency we included the interactions $V^{ee}_r$ up to 
range $d_U =3$.
In Fig. \ref{fig:EffJU}(b) we show the interaction potential $U_r$ 
of Eq.~(\ref{eq:UrAtdef}), i.e., Eqs.~(\ref{eqs:UrAt}) without $V^{ee}_r$. 
Clearly, the nearest-neighbor interaction $U_1$ is stronger than 
$U_{r \geq 2}$, neglecting which would correspond to the spin-lattice 
model studied in the text.  
We can now write the interaction term of the Hamiltonian as 
\begin{equation}
\mathcal{H}^{(2)}_U = \sum_{K} \ket{K}\bra{K} 
\otimes \sum_{r=1}^3 U_{r} \ket{r}\bra{r} , 
\label{eq:2HUat}
\end{equation}
which has the same form as Eq.~(\ref{eq:app:2HU}). 

To summarize, the total Hamiltonian for two $\ket{e}$ excitations 
in a lattice of Rydberg dressed atoms is
\begin{equation}
\mathcal{H}^{(2)} = \mathcal{H}_J^{(2)} + \mathcal{H}_U^{(2)}, \label{eq:2AtFullH}
\end{equation}
where $\mathcal{H}_J^{(2)}$ and $\mathcal{H}_U^{(2)}$ are given 
by Eqs.~(\ref{eq:2AtExcKr}) and (\ref{eq:2HUat}), respectively. 
In Fig.~\ref{fig:BoundPairSpectrum}(a) of the main text we show the spectrum 
of this Hamiltonian. The scattering states are insensitive to the variations 
of $J_d(r)$ and $U_r$ at short range $r \leq 3$, so the scattering
spectrum is well reproduced by the spin-lattice model Hamiltonian 
with $r$-independent hopping rates $J_d$ and only the nearest-neighbor 
interaction $U_1$. The spin-lattice model approximates well also the 
bound states of Hamiltonian~(\ref{eq:2AtFullH}), especially for $U_r$ 
dominated by the nearest-neighbor interatomic interaction $V^{ee}_1$ and 
constant $J_1$ achieved for $V^{se}_1 = -D_2$, which is used in 
Fig.~\ref{fig:BoundPairSpectrum}(a). 
 
\begin{figure}[t]
  \centerline{\includegraphics[width=0.85\columnwidth]{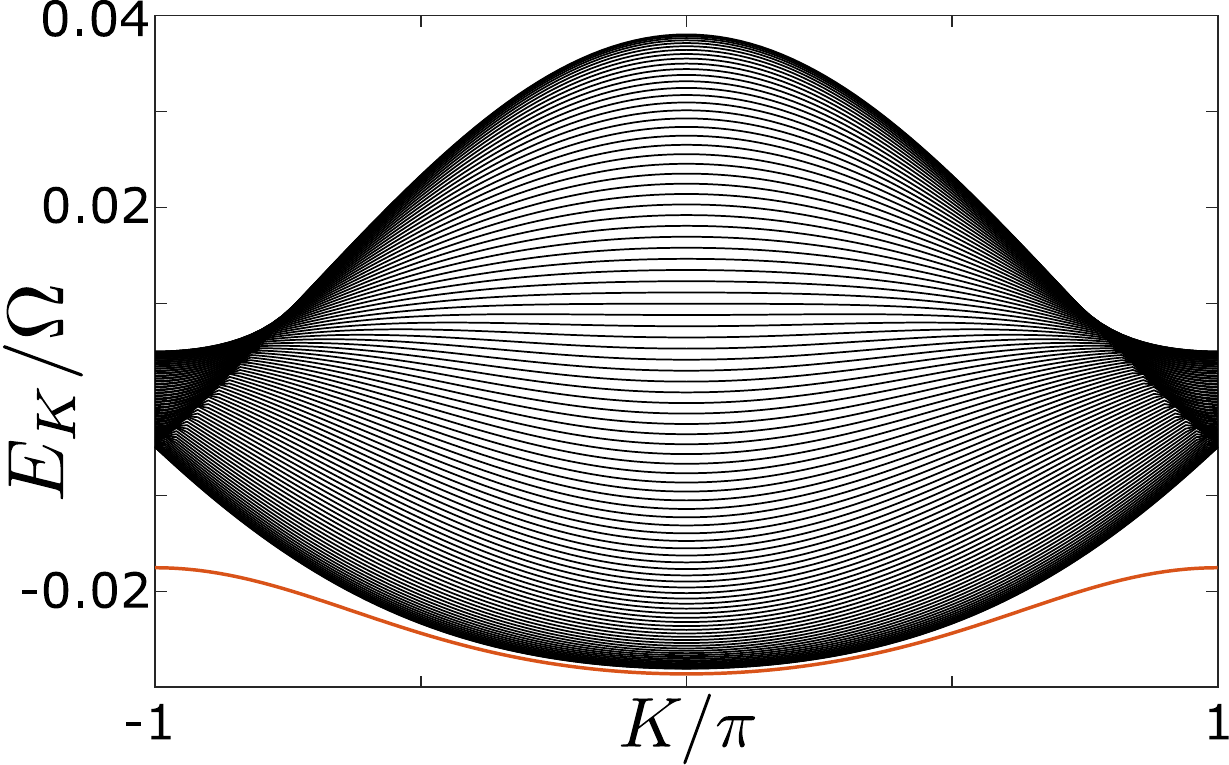}}
  \caption{Scattering and bound spectrum for two $\ket{e}$ excitations 
    in a lattice, with the parameters as in Fig.~\ref{fig:EffJU}. }
  \label{fig:BoundByDressing}
\end{figure}

Note that even without interatomic interactions, $V^{ee}_r,V^{se}_r = 0$, 
we still have non-vanishing effective interaction 
$U_1 \simeq 2 \frac{\Omega^2 D_1^2}{\Delta^3}$, which is, however, 
too weak compared to $J_1 \simeq \frac{\Omega^2 D_1}{\Delta^2}$ 
to sustain a bound state, see the lower panel of 
Fig.~\ref{fig:ExistenceOfBoundStates} in the main text. 
But strong enough interatomic interaction 
$|V^{se}_1| \simeq \Omega, D_1 \ll \Delta$ 
resulting in $U_1 \simeq - 2 \frac{\Omega^2 V^{se}_1}{\Delta^2}$ can sustain 
two-excitation bound state, as shown in Fig.~\ref{fig:BoundByDressing}. 
The corresponding hopping rate $J_1(r)$ has now sizable $r$-dependence, 
see Fig.~\ref{fig:EffJU}.

\paragraph{Experimental considerations.}

A suitable system to realize the spin lattice model and observe 
the bound states of Rydberg (spin) excitations is a defect-free
chain of cold atoms in a one-dimensional optical lattice potential 
or an array of microtraps \cite{Barredo1021,Endres2016}.
The microtraps can be spaced by $a = 5-20\:\mu$m, and each microtrap 
confines the ground state atom within $\Delta a \simeq 1\:\mu$m.  
We take the atomic parameters similar to those in the recent
experiments \cite{Begin2013,Barredo2015,Browaeys2016}. 
The ground state of Rb atoms $\ket{g} = 5S_{1/2}$ can be dressed
with the Rydberg state $\ket{s} = 63P_{1/2}$ by a detuned UV laser
with the Rabi frequency $\Omega/(2 \pi) \simeq 5\:$MHz and detuning
$\Delta/(2 \pi) \simeq 33\:$MHz ($\Omega/\Delta = 0.15$). 
The excited Rydberg state $\ket{e} = 62D_{3/2}$ can be populated 
by a two photon transition from the ground state using laser 
beams focused onto the desired atoms. 
With the above Rydberg states $\ket{e}$ and $\ket{s}$, 
the dipole-dipole coefficient for the exchange interaction 
$D = C_3/r^3$ is $C_3 = 7950\:$MHz$\:\mu$m$^3$ and the 
van der Waals coefficient for the interaction $V^{ee} = C_6/r^6$ 
is $C_6 = 730\:$GHz$\:\mu$m$^6$ \cite{Begin2013,Barredo2015,Browaeys2016}.
The lifetime of state $\ket{e}$ is $\tau_e \simeq 100\:\mu$s, and the 
dressing state $\ket{s}$ has a similar lifetime $\tau_s \simeq 135\:\mu$s 
but its decay is suppressed by the factor of $\Omega^2/\Delta^2$. 

With the lattice constant $a \simeq 10\:\mu$m, we have 
$U_1 \simeq 730\:$kHz $J_1 \simeq 180\:$kHz and $J_2 \simeq 22\:$kHz.
The hopping rates are larger than the Rydberg state decay rates,
which permits observation of coherent dynamics of the Rydberg excitations
in the lattice. At the same time, the interaction $U_1 = 4 J_1$ will
support strongly bound states of Rydberg excitations.

The dressed ground state atoms are tightly confined by the microtraps,
but the atoms in the Rydberg state are usually not trapped.  
During the interaction, the Rydberg excited atoms experience 
a repulsive (or attractive, if $C_6 <0$) force 
$F = - \partial_r V^{ee}(r) = -6C_6/r^7$ which can result 
in their displacement $\Delta r$ from the equilibrium lattice positions. 
We can estimate the displacement for a pair of atoms at the neighboring
lattice sites, $r=a$, as $\Delta r \simeq \frac{F(a)}{2m} t^2$, where
$m$ is the atomic mass and $t \simeq J_{1,2}^{-1}$ is the timescale  
of the interaction. We then obtain $\Delta r = 3-200\:$nm, which is 
still smaller than the trap waist $\Delta a$.  

We finally note that similar parameters of the spin lattice model 
can be obtained for atoms in optical lattices with a smaller 
period $a \lesssim 1\:\mu$m by choosing lower-lying Rydberg states
$\ket{s}$ and $\ket{e}$. Such states, however, have shorter lifetimes,
which necessitates larger hopping rates $J_d$ obtained with stronger
dressing lasers. Furthermore, at small interatomic separation, the 
van der Waals interactions between the untrapped Rydberg-excited atoms 
will exert stronger force, leading to the displacement of atoms 
comparable to the lattice spacing. This can be mitigated by using
``magic wavelength'' optical lattices which simultaneously trap the 
atoms both in the ground state and the Rydberg state \cite{Zhang2011}.

%

%

\end{document}